\documentclass[10pt,twocolumn,showpacs,preprintnumbers,amsmath,amssymb,aps,prb,
superscriptaddress,floatfix,footinbib]{revtex4-1}

\usepackage{graphicx}
\usepackage{amsmath}
\usepackage{amssymb}
\usepackage{hyperref}

\usepackage[active]{srcltx}
\usepackage{lipsum}
\usepackage{color}

\newif\iftransport
\transporttrue
\newif\ifMPG
\MPGfalse
\newif\iftodo
\todotrue

\usepackage[usenames,dvipsnames,svgnames,table]{xcolor}

\begin{document}


\title{Spin and orbital Hall effect in non-magnetic transition metals:
  extrinsic vs intrinsic contributions.} 
\author{Sergiy Mankovsky}
\affiliation{%
  Department  Chemie,  Physikalische  Chemie,  Universit\"at  M\"unchen,
  Butenandtstr.  5-13, 81377 M\"unchen, Germany\\}
\author{Hubert Ebert}
\affiliation{%
  Department  Chemie,  Physikalische  Chemie,  Universit\"at  M\"unchen,
  Butenandtstr. 5-13, 81377 M\"unchen, Germany\\}


\begin{abstract}
  Kubo's linear response formalism has been used to calculate the orbital
  Hall conductivity (OHC) for non-magnetic undoped and doped transition metal systems,
  focusing on the impact of different types of disorder and the role of
  vertex corrections for the OHC. The doping- and
  temperature-dependence of the OH conductivity have been
  investigated and compared with corresponding results for the spin Hall
   conductivity (SHC).  
 A strong difference has been found between the results for undoped and
 doped metallic systems. For elemental systems at finite temperature a
 dominating role of the intrinsic contribution to 
the temperature-dependent OH and SH conductivities is found.  
 Moreover, the different temperature dependent behavior of
 the intrinsic SOC-independent OHC and SOC-driven SCH indicates
 a non-trivial relationship between these quantities.
  It is shown, that in contrast to the intrinsic part of
  the OH and SH conductivities, the extrinsic contributions in doped
  systems are determined by spin-orbit coupling for both of
  them. It is dominating at low temperature, strongly decreasing
  at higher temperatures due to the increasing impact of the electron-phonon
  scattering.
\end{abstract}

\keywords{Suggested keywords} 
\maketitle

\section{Introduction}

The occurrence of a spin current in solids in the direction
perpendicular to an applied electric field, called spin Hall effect
(SHE), is a well established phenomenon studied both experimentally and 
theoretically. This effect is broadly discussed 
in the literature concerning its origin, showing in particular that its
intrinsic contribution is determined by topological features of the
electronic band structure, while the extrinsic part is seen as a
consequence of asymmetric electron scattering in the presence of
different types of disorder in the material (see, e.g., Ref.\
[\onlinecite{SVW+15}] and references therein).  
Another type of Hall effect, called orbital Hall effect (OHE), 
predicted for doped semiconductors \cite{BHZ05} and for transition metals
\cite{KTH+09}, implies for a finite sample an angular momentum
accumulation due to the angular momentum flow in the direction
perpendicular to applied electric 
field. The origin of the OHE is still under discussion.
In contrast to the spin Hall conductivity (SHC), representing the SHE,
the intrinsic orbital Hall conductivity (OHC) connected to the OHE is
almost independent of the spin-orbit coupling (SOC). Moreover, the sign
of the OHC can differ from that of the SHC
\cite{KTH+09,GLO+24}. Tanaka et al. \cite{TKN+08} and H. Kontani et
al. \cite{KTH+09} concluded that the OHC 
has to be always positive for $4d$ and $5d$ transition metals. 
On the other hand, recent findings by Go et al. \cite{GLO+24}
demonstrate, that the intrinsic OHC can have a negative sign
for some late  $3d$, $4d$ and $5d$ elements.
Discussing the origin of the SHE and OHE,
Kontani et al. \cite{KTH+09} have shown that the SHE in
transition metals may be seen as a SOC-driven spin flow accompanying      
the transverse flow of orbital angular momentum in the presence of 
an applied electric field.
 As a consequence, the SHC may be 
approximated by the product of the OHC and the 
spin polarization at the Fermi energy, arising due to the SOC
\cite{KTH+09}.  
This explains in particular, the reason, why the intrinsic OHC
  exceeds the SHC, as well as why they have different sign for some
  materials.

Coming to the origin of the intrinsic OHE, the transverse orbital
angular momentum flow induced by an applied electric field may be seen
as a consequence of the orbital texture in momentum space\cite{HLK23}, even
if the orbital angular momentum at each atomic site 
$\vec{l}_i$ is completely quenched in equilibrium. 
Thus, Go et al. \cite{GJKL18} suggested a mechanism giving rise
for the OHC, following an idea of the Sinova et al. \cite{SCN+04}
to explain the origin of the intrinsic SHE. The latter mechanism implies a
spin current arising due to the spin texture in momentum space,
which is induced by spin orbit coupling (SOC) and therefore is seen as a 
relativistic effect in contrast to the OHE. According to the suggestion
by Go et al. \cite{GJKL18}, 
a finite k-dependent orbital angular momentum in reciprocal space
can be dynamically induced without SOC but due to a field-induced
excitation of the electronic structure, leading at the end to a 
non-vanishing orbital current.
A giant orbital Hall effect for a broad
class of centrosymmetric materials was reported by various 
authors on the basis of theoretical investigations 
\cite{KTH+08,TKN+08,SBNO18,GJL+21,GLO+24}.
Furthermore, several theoretical predictions have been confirmed also 
experimentally. In particular, an electric field induced orbital angular
momentum accumulation has been observed recently also experimentally in
non-magnetic \cite{CJK+23} as well as in magnetic materials \cite{LAB+23}.

In the case of doped materials and random alloys, the conductivity tensor
will be modified due to the impact of disorder. 
For instance, a modification of the SHE due to disorder may occur, on the one
hand side, as a result of corresponding modification of the electronic
structure having an impact on the intrinsic part of the SHE.
On the other hand, the SOC-driven asymmetric scattering of the charge
carriers with opposite spin (skew scattering and side-jump
scattering) leads to the extrinsic contribution 
which may strongly modify the SH conductivity (see, e.g. \cite{CB01,IBM04,SVW+15}).

In the case of the OHE, only few results have been reported in the 
literature so far, which concern the impact of disorder on the OHC.
In the case of $p$-doped silicon studied using the Luttinger model, it was
shown that the vertex corrections due to impurity scattering vanish
identically, and the OHE is represented essentially by the intrinsic
contribution \cite{BHZ05}. On the other hand, Tanaka et al.\cite{TKN+08}
reported about a finite correction to the OHC calculated for transition
metals, which is introduced by disorder.

Tang and Bauer \cite{TB24} have also investigated the role of disorder
for the OHE solving the quantum Boltzmann equation. They have shown that
the corrections to the OHC due to disorder are very sensitive to the
scattering potential and pointed out that the predictions by Bernevig et 
al.\cite{BHZ05} are valid only in the limit of a short-range scattering
potential.
Using a finite-range scattering potential, Tang and Bauer calculated 
the correction to the OHE due to diffuse scattering which was taken into 
account via a scattering-in contribution to the collision term in the
Boltzmann equation. These calculations already demonstrated the important
 role of disorder for the OHC.
Based also on the findings of previous works (see, e.g. Refs. [\onlinecite{IBM04,TKN+08}]), which concern the
role of the scattering potentialfor the SHE and OHE, one can conclude
that first principles calculations of the OHC are of great importance to get
reliable results on the extrinsic contribution to this quantity. 
A strong impact of disorder on the OHC has been reported recently by
Liu and Culcer \cite{LC24} who investigated the OHE in graphene  and
transition metal dichalcogenides.  
They found that the extrinsic OHC in doped systems provides
the dominant contribution to the OHE in the case when the Fermi level
crosses the conduction or valence bands. The authors attribute this
extrinsic effect to the skew and side jump scattering, making use of
the analogy with the SHE in the investigated materials.
One has to note in addition, that there is no explicit discussion in
the literature so far concerning the scattering mechanisms
responsible for the asymmetric scattering of the electrons with
different orbital angular momentum. Expecting that the scattering
asymmetry of the orbital and spin angular momentum are closely connected
to each other due to SOC, we will discuss and compare the various scattering
mechanisms both for the OHE and SHE.

Below, we will present the results of first principles calculations for
the OHC and SHC. We will compare the properties of the OHC and SHC for pure
materials, calculated for finite temperature, discussing both
contributions, extrinsic and intrinsic, and demonstrate how these
contributions depend on temperature. Next, we will deal with doped
transition metal systems, and  will discuss the
concentration and temperature dependence properties of the
extrinsic and intrinsic OHC in comparison with the SHC. Moreover, we will  
demonstrate the crucial role of spin-orbit interaction on the extrinsic
contribution both for the OHE and  the SHE.

\section{Theoretical details} \label{Theoretical details}

Here we will describe some technical details relevant for the
discussions below on the results of our calculations.
As the calculations of the OH and SH conductivities are closely
connected to each other, all discussions of their properties 
will be done in parallel. This takes into account, that the properties
of the SHE were already previously investigated and discussed
extensively in the literature \cite{SCN+04,Hof13,SVW+15,ZZD+24}. 

The calculations of the OHC, similar to the SHC
\cite{Low10,LKE10,LGK+10,LGK+11}, are based on Kubo's linear 
response formalism using the so-called Bastin formula \cite{BLBN71} (for
more details see the Appendix).
The underlying electronic structure is calculated making use of the fully
relativistic KKR Green's function method\cite{EBKM16,SPR-KKR8.5}. The OH and SH 
conductivity tensors are calculated accounting for the vertex corrections
\cite{But85,LGK+10,LGK+11} to describe properly the corresponding transport  
properties in the presence of various types of disorder.
Formally a corresponding conductivity tensor element may 
be written as follows \cite{Low10} (omitting the angular
momentum indices): 
\begin{eqnarray}
\sigma_{\mu\nu}^{\rm VC} &\propto& \, Tr {J}_{\mu}(z_2,z_1)
                               \left[(1-\chi\,w)^{-1}\chi\right]
                                {j}_{\nu}(z_1,z_2) \,.
\end{eqnarray}
The current density operator $\vec{j}$ representing the perturbation
$j_\nu$ with corresponding matrix elements is given in 
relativistic form by $\vec{j} = ec \vec{\underline{\alpha}}$ with the velocity operator
$\vec{v} = c\,\vec{\underline{\alpha}}$ and $\underline{\alpha}_\mu$ the standard Dirac
matrices. For calculations of the SHC, the response   
function is given by the spin current density operator $\hat{J}^{{\rm s},\beta}_\mu$
\cite{LKE10,LGK+11}. The OHC is calculated using for the response orbital
current density the operator given by the expression 
 $\hat{J}^{{\rm o},\beta}_\mu = \frac{1}{2} c (\hat{l}^{\beta} \underline{\alpha}_\mu + 
 \underline{\alpha}_\mu \hat{l}^{\beta})$, where $\hat{\vec{l}}$ is the
 orbital angular momentum operator. Below we will use the shorthand
 notation   
 $\sigma_{\rm SH}$ and $\sigma_{\rm OH}$ for the tensor elements
 $\sigma^{{\rm (S)},y}_{xz}$ and $\sigma^{{\rm (O)},y}_{xz}$ of the 
 spin polarization and orbital polarization conductivity tensors,
 respectively. 
 
As it was discussed in the literature (see, e.g.,
Refs.\ [\onlinecite{CB01,SVW+15,LGK+11}], the extrinsic SHE  is 
associated with the vertex corrections to the expression for 
the conductivity tensor, which account for the SOC-driven side-jump and 
skew scattering in the electron transport. 
Expanding the term $(1-\chi\,w)^{-1}$ accounting
for vertex corrections in a Taylor series, leads to the expression
\begin{equation} \label{Eq_1}
\sigma_{\mu\nu}^{\rm VC} \propto \sigma_{\mu\nu}^{\rm NVC} +
{J}_{\mu}(z_2,z_1) \left[   \chi\,w\,\chi
  +\chi\,w\,\chi\,w\,\chi + ...   \right]
{j}_{\nu}(z_1,z_2) \; . 
\end{equation}
from which one can see that the difference $\sigma_{\mu\nu}^{\rm VC} -
\sigma_{\mu\nu}^{\rm NVC}$ characterizes the extrinsic contribution to the
SHC. Accordingly, the intrinsic part of the SHE may be 
obtained by performing the calculations without vertex corrections.
The same refers also to the OHC. It is important to remind also,
that the vertex corrections, giving the extrinsic contribution to the
SHC, are associated with the Fermi surface electrons\cite{OSN08}.
In the superclean regime (i.e. with the impurity concentration $x \to
0$) the scattering is dominated by the skew-scattering 
mechanism \cite{OSN06} and the SHC is linearly connected with the
electrical longitudinal conductivity $\sigma_{\mu\mu}$ according to
$\sigma^{\rm(S) skew}_{\mu\nu} = S\,\sigma_{\mu\mu}$, where $S$ is 
the so-called skewness factor. This implies, that 
  $\sigma^{\rm skew}_{\mu\nu}$ diverges 
  together with $\sigma_{\mu\mu}$ as $1/x$ in the pure limit.
In contrast to the skew scattering, the side jump and intrinsic contributions in
the low-concentration regime are often comparable in magnitude and
weakly depend on impurity content \cite{Sin08,LGK+11,CWEK15}. As a result, the
SHC can be represented by the expression  
\begin{equation} \label{Eq_2}
  \sigma_{\mu\nu}^{{\rm (S)}}  = \sigma_{\mu\nu}^{{\rm (S) intrinsic}} +
  \sigma_{\mu\nu}^{{\rm (S) sj}} + S\sigma_{\mu\mu}  \; . 
\end{equation}
Assuming that the same scattering mechanisms are responsible for the
extrinsic OHC as well, we will discuss below the skew scattering and side jump
contributions also for the OHC.

In order to investigate the impact of the SOC on the SHC and OHC,
corresponding calculations have been performed applying a scaling of the 
speed of light $c$ allowing a continuous transition from the relativistic
description of transport properties (i.e., without scaling $c = c_0$) to
the non-relativistic limit corresponding to $c = \infty$. As the leading
relativistic corrections to the Schr\"odinger Hamiltonian vary as
$1/c^{2}$, we will use the scaling parameter $\xi = 
\big(\frac{c_0}{c}\big)^2$ \cite{BVE96}. It will be referred to in the
discussions below as the SOC scaling factor, assuming that among  all the
relativistic effects, SOC has the dominating impact on the SHC and OHC.

The calculations have been performed using the spin-polarized
relativistic KKR  Green function  method
\cite{SPR-KKR8.5,EKM11} in combination with the atomic sphere
approximation (ASA).
The exchange-correlation potential was calculated within the local spin
density approximation (LSDA) to spin density 
functional theory (SDFT), using a parametrization as given by Vosko et al.
\cite{VWN80}. The angular momentum expansion of the Green function was
 cut-off at $l_{\rm max} = 3$.
A $k$-mesh including up to $10^8$ grid points in the full Brillouin zone
(BZ) was used for the integration over the BZ when calculating the  
conductivity tensors.

The temperature dependent properties of the OHC and SHC have been
investigated by accounting for the impact of thermally induced lattice
vibrations, assuming that the effects of the temperature dependent
Fermi-Dirac distribution function can be neglected in the  temperature 
regime under consideration.
These calculations have been performed making use of the alloy analogy
model \cite{EMC+15} based on the adiabatic approximation, that implies that a discrete
set of $N_{v}$ displacement vectors $\Delta \vec{R}^q_v(T)$ with
probability $x^q_v$ ($v=1,..,N_{v}$) is constructed for each basis atom $q$
within the crystallographic unit cell.
The vectors  $\Delta \vec{R}^q_v(T)$
 are connected with the temperature
dependent root mean square displacement $(\langle u^2\rangle_T)^{1/2}$
according to the relation:
%
\begin{equation}
\label{eq:displacement}
\sum_{v=1}^{N_{v}} {x^q_{v}} | \Delta \vec{R}^q_v(T) |^2 = \langle u_q^2\rangle_T \;.
\end{equation}
%
For the applications presented below, 
the temperature dependent
 root mean square displacement is estimated using  Debye's theory,
 providing a simple connection between  $\langle
u_q^2\rangle_T$ and the lattice temperature.

\section{Results}

 In this section we will consider two types of non-magnetic materials,
 undoped and doped. In the latter case we will show the impact of
 chemical disorder on the OHE discussing it in parallel with the SHE.
 Furthermore, for both types of materials we will discuss the
 temperature dependence of the OHC and SHC.

\subsection{Elemental transition metal materials at finite temperature.}

\begin{figure}
 \begin{center}
 \includegraphics[angle=0,width=0.8\linewidth,clip]{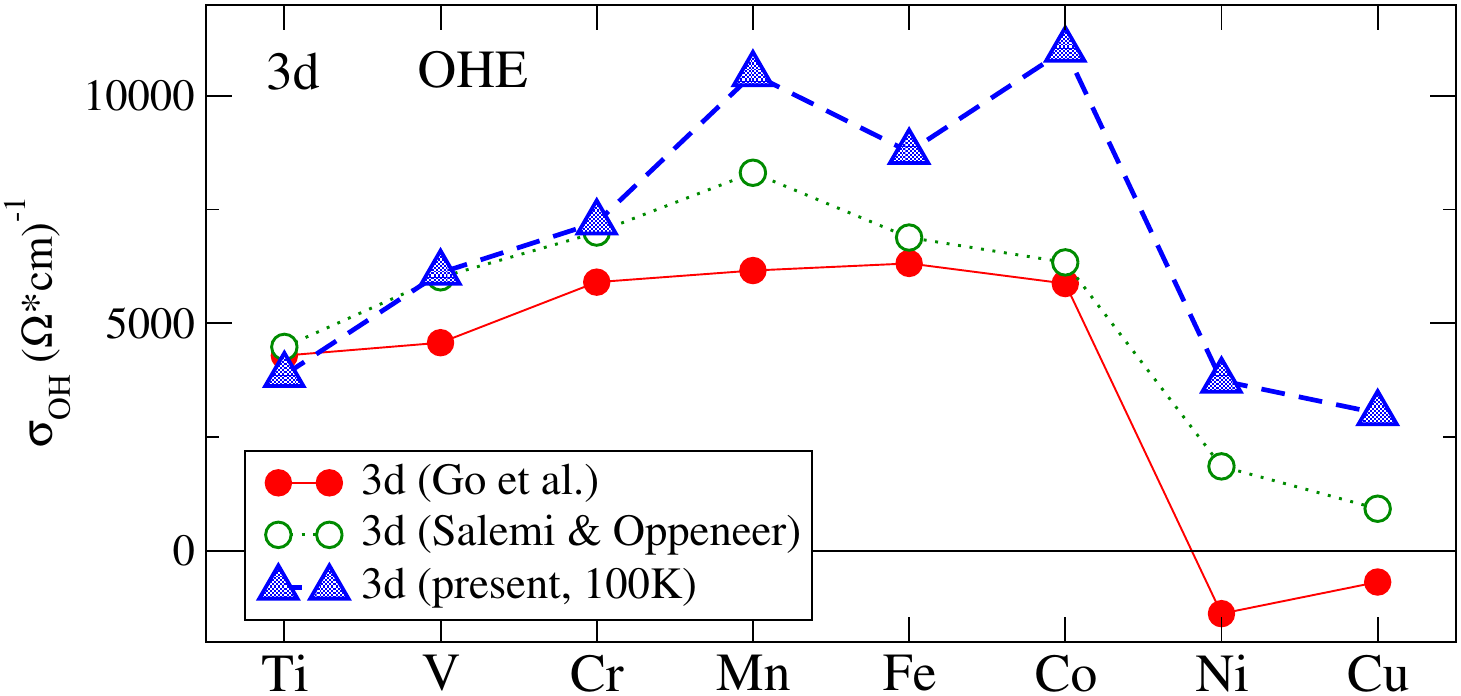}
 \includegraphics[angle=0,width=0.8\linewidth,clip]{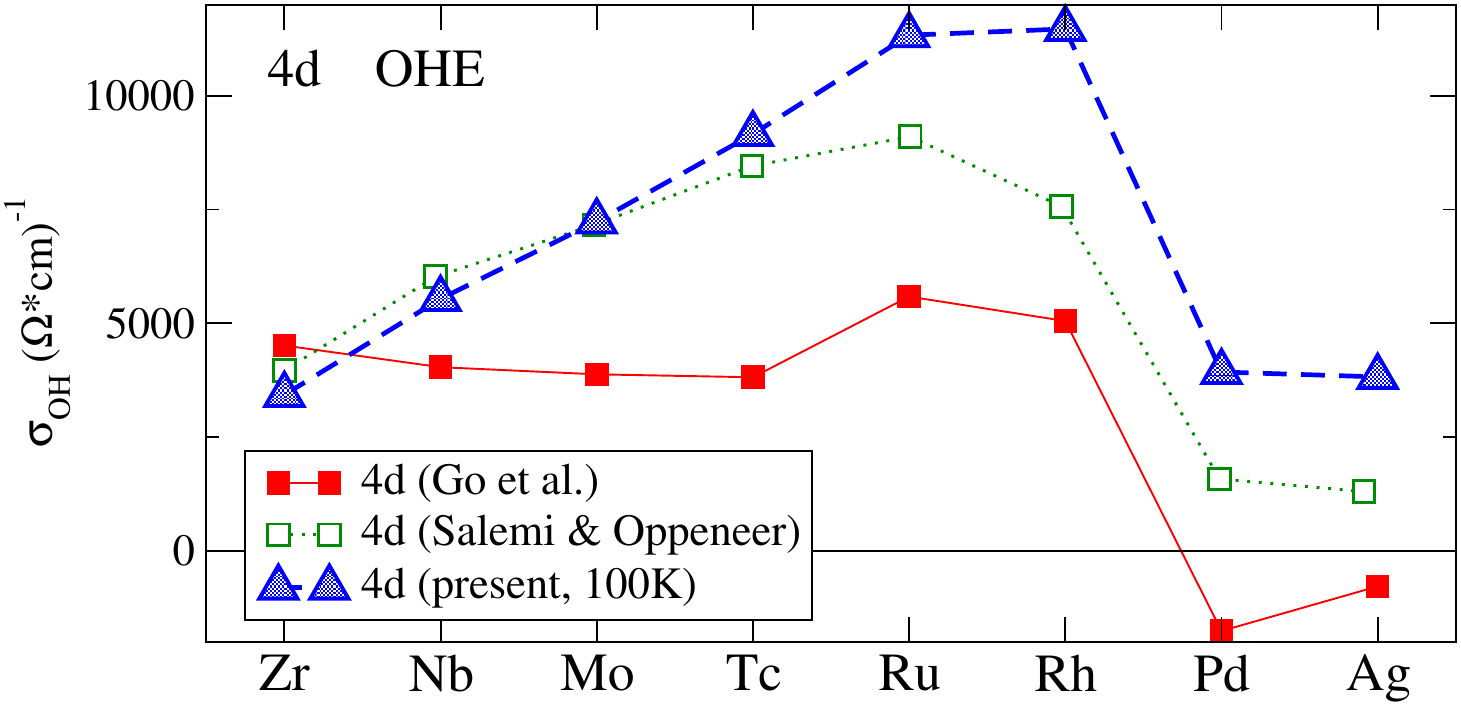}
 \includegraphics[angle=0,width=0.8\linewidth,clip]{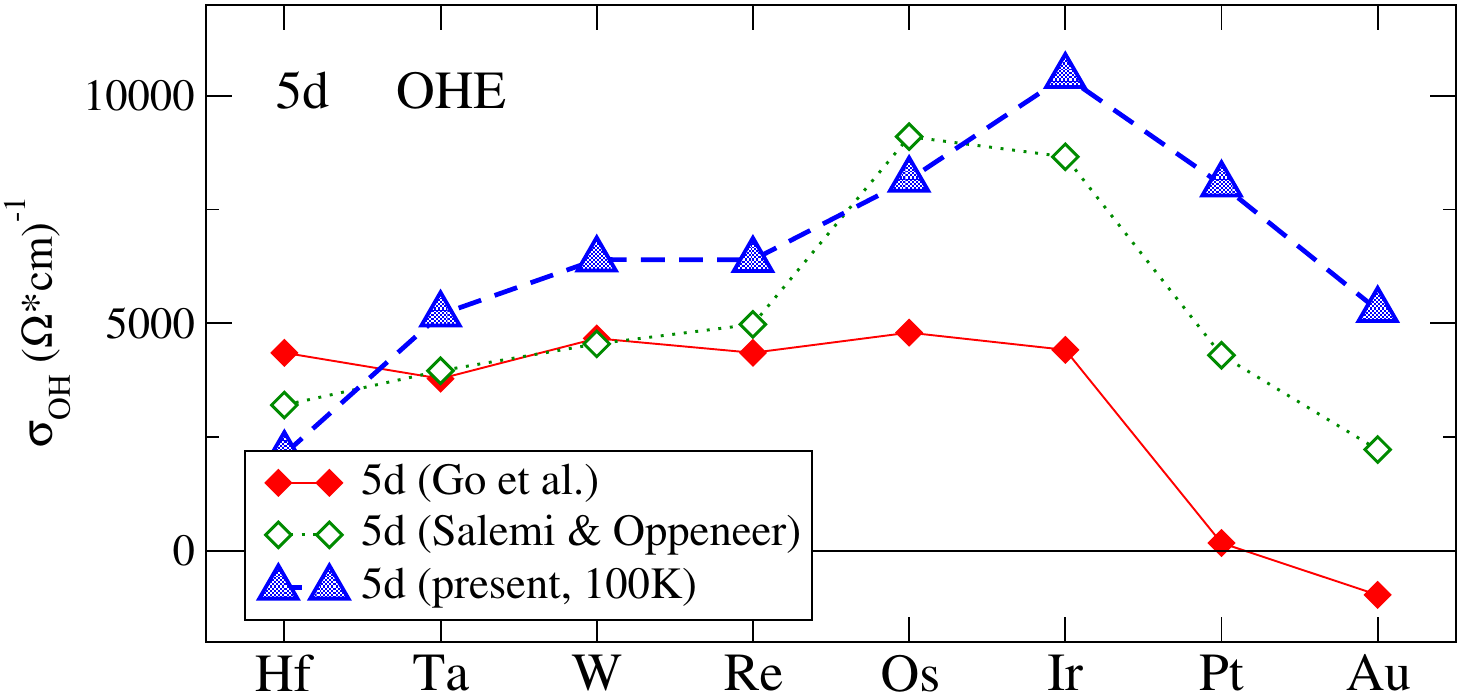}
  \caption{\label{Pure_OHE-vs-litera} Orbital Hall conductivity $\sigma_{\rm OH}$
    for the elemental paramagnetic $3d$, $4d$ and $5d$ materials,
    calculated for $T = 100$K in comparison with the results by Go et
    al. \cite{GLO+24} and Salemi and Oppeneer \cite{SO22} for $T = 0$\,K. }
\end{center}
\end{figure}

\begin{figure}
 \begin{center}
 \includegraphics[angle=0,width=0.8\linewidth,clip]{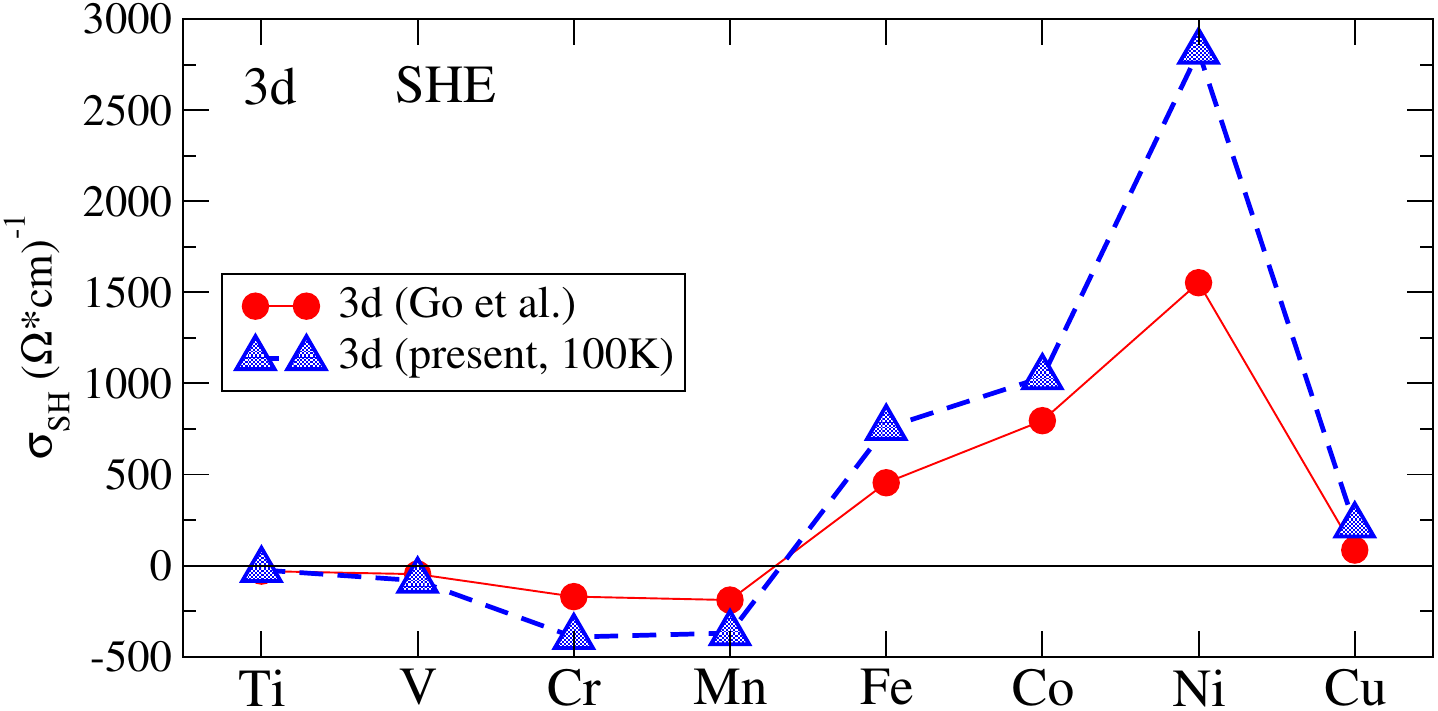}
 \includegraphics[angle=0,width=0.8\linewidth,clip]{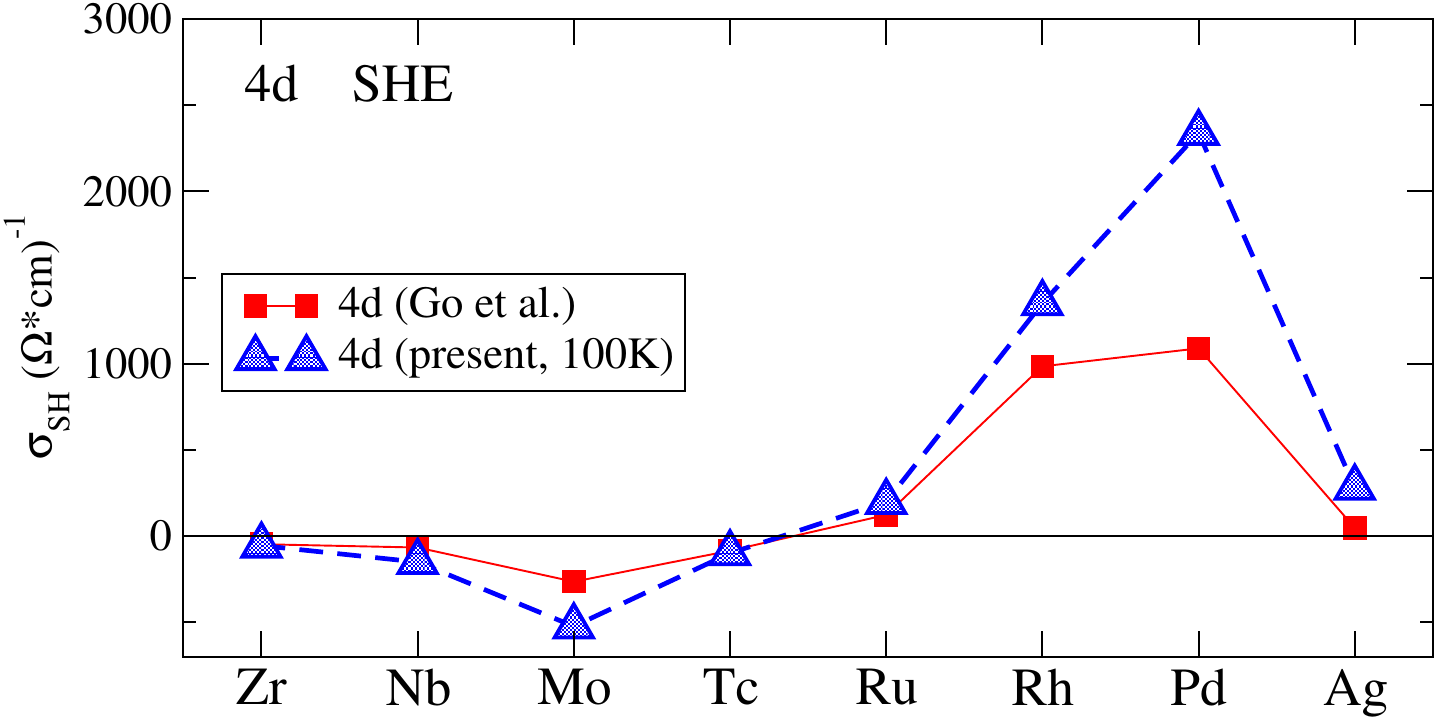}
 \includegraphics[angle=0,width=0.8\linewidth,clip]{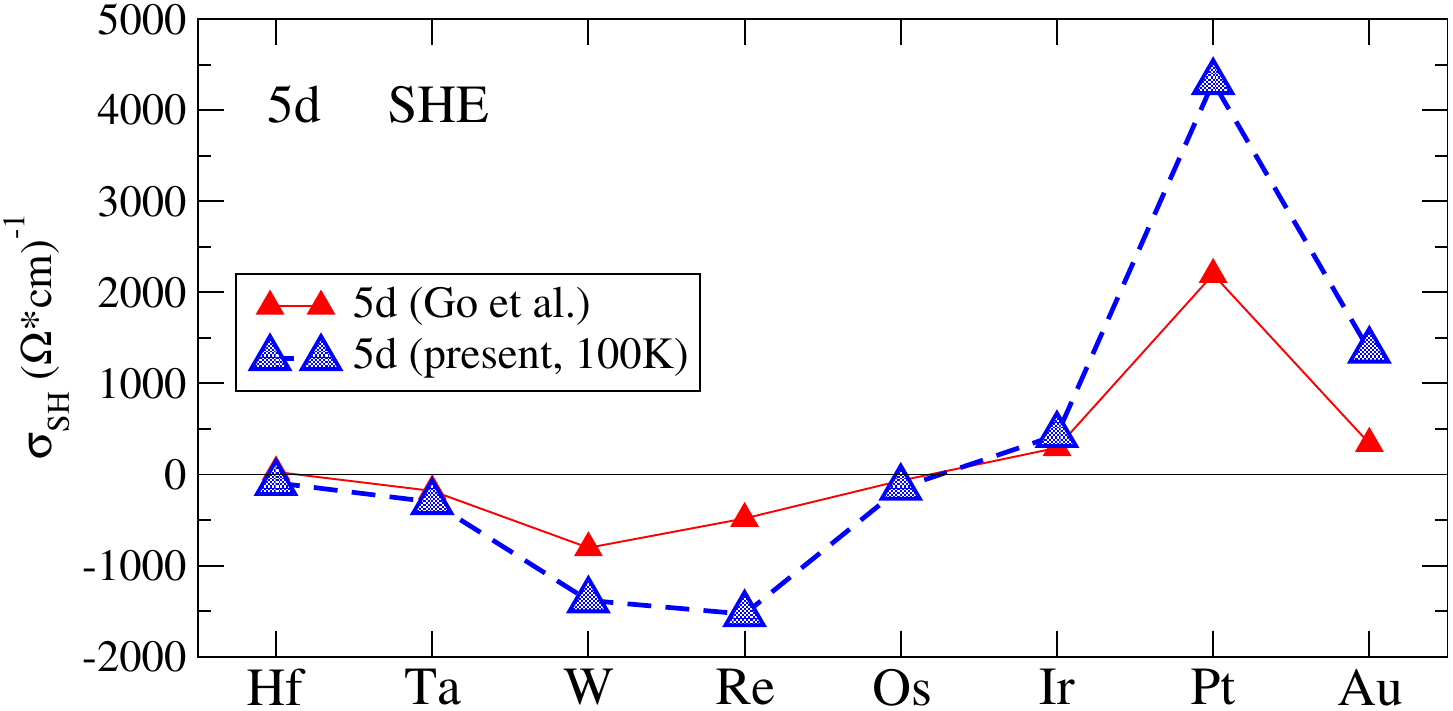}
  \caption{\label{Pure_SHE-vs-litera} Spin Hall conductivity
    $\sigma_{\rm SH}$ for the elemental paramagnetic $3d$, $4d$ and $5d$
    materials, calculated for $T =  100$K in comparison with the results by Go et
    al. \cite{GLO+24} for $T = 0$\,K. }
\end{center}
\end{figure}

First, we compare our results for elemental  $3d$,  $4d$
and $5d$ transition metals with some first-principles results
available in the literature. This comparison can be seen in
Figs. \ref{Pure_OHE-vs-litera} and  \ref{Pure_SHE-vs-litera} for the OHC
and SHC, respectively. 
For numerical reason the present calculations were performed for the lattice
temperature $T = 100$\,K. 
The temperature was taken low enough to avoid a
noteworthy temperature dependent extrinsic
contribution to the Hall effects caused by the electron scattering by
lattice vibrations. This ensures that the obtained results can be
directly compared with the results by other groups giving the
intrinsic OHC and SHC \cite{GLO+24,SBNO18} for $T = 0$\,K. One can see
a reasonably good 
agreement between the various results, with the difference partly
attributable to the finite temperature assumed for the calculations in
the present case. 

\begin{figure*}
 \begin{center}
 \includegraphics[angle=0,width=0.22\linewidth,clip]{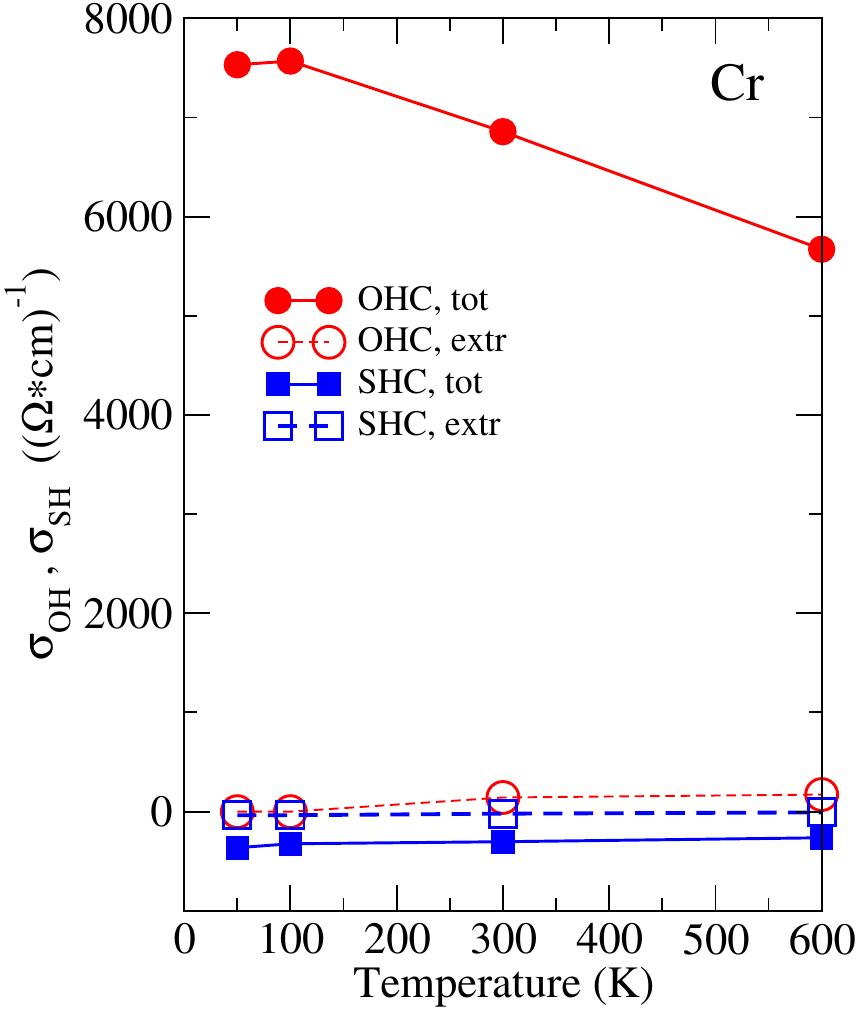}
 \includegraphics[angle=0,width=0.22\linewidth,clip]{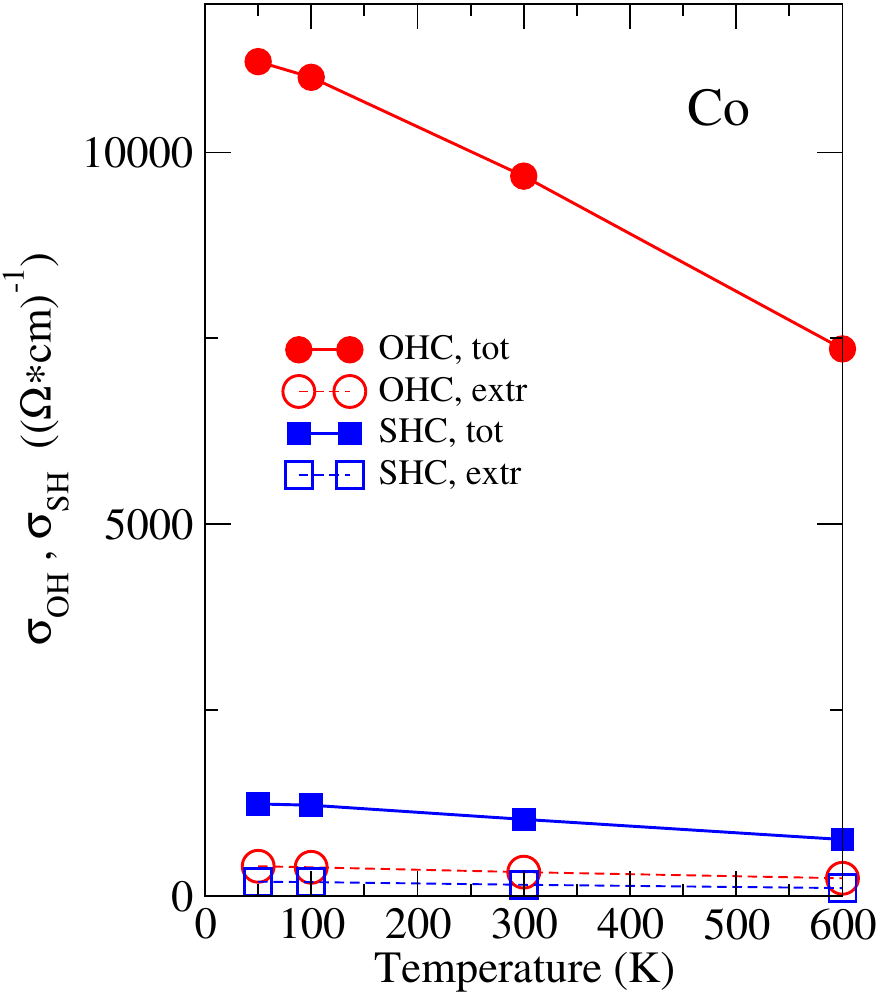}
 \includegraphics[angle=0,width=0.22\linewidth,clip]{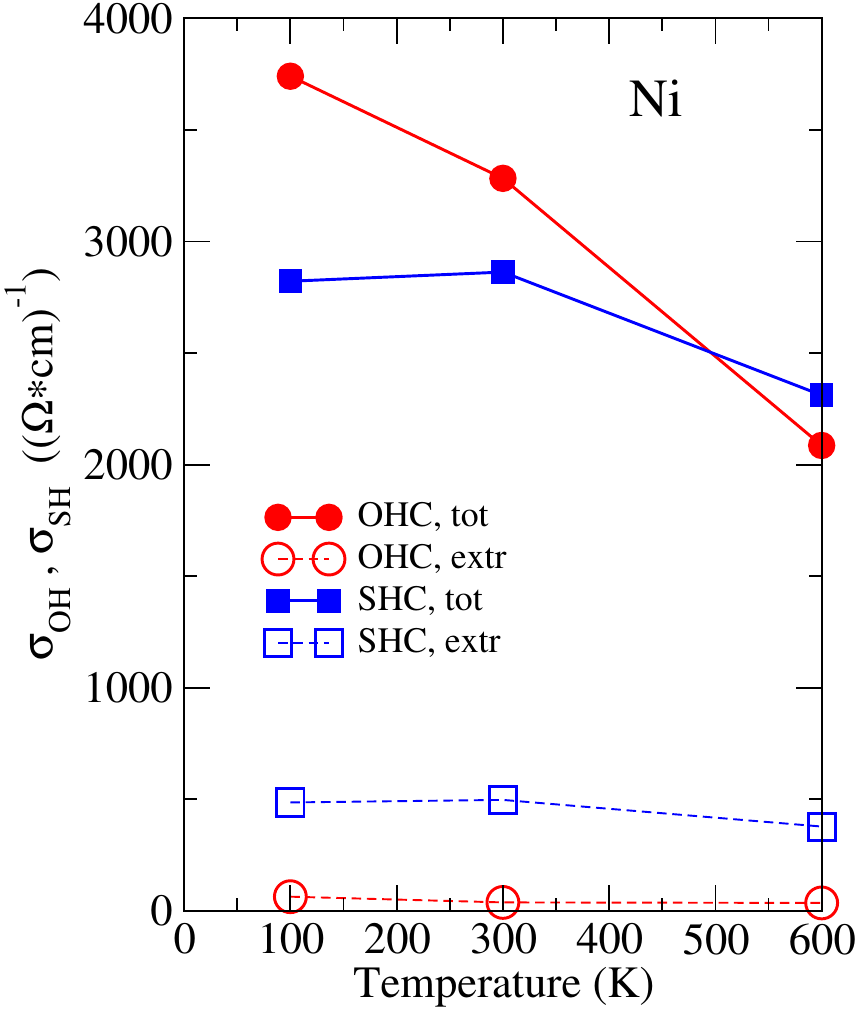}
 \includegraphics[angle=0,width=0.22\linewidth,clip]{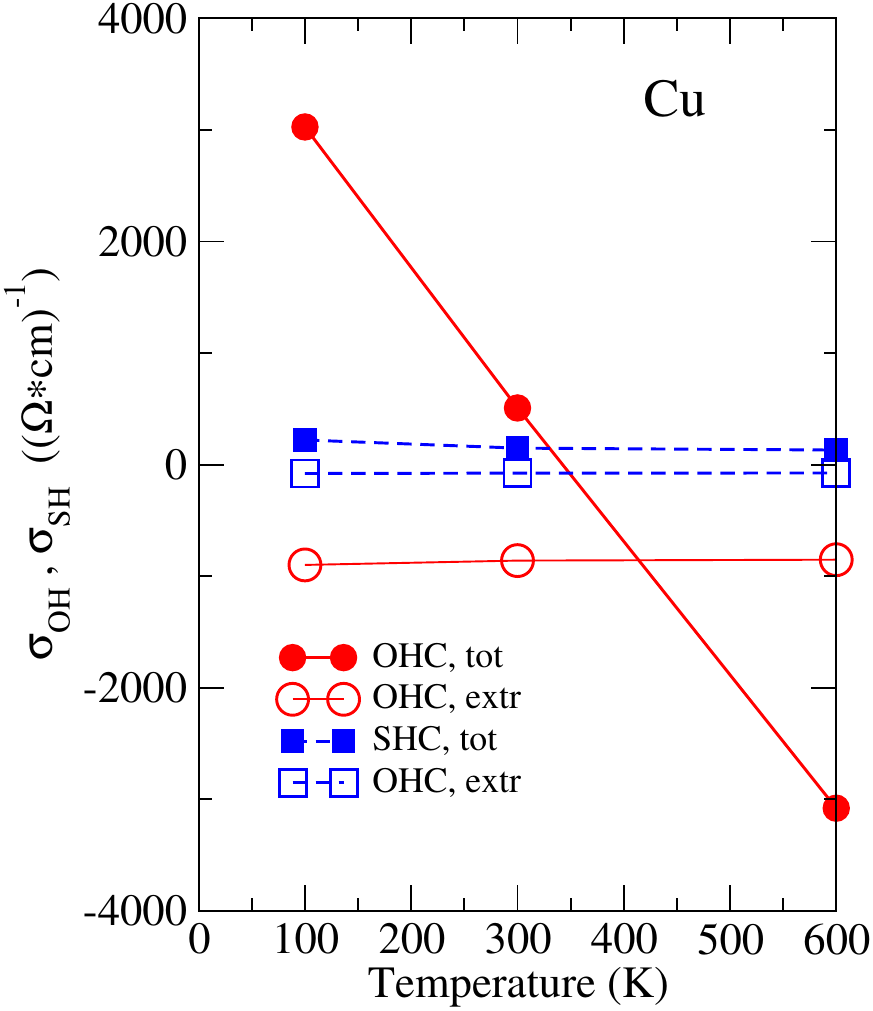}
 \includegraphics[angle=0,width=0.22\linewidth,clip]{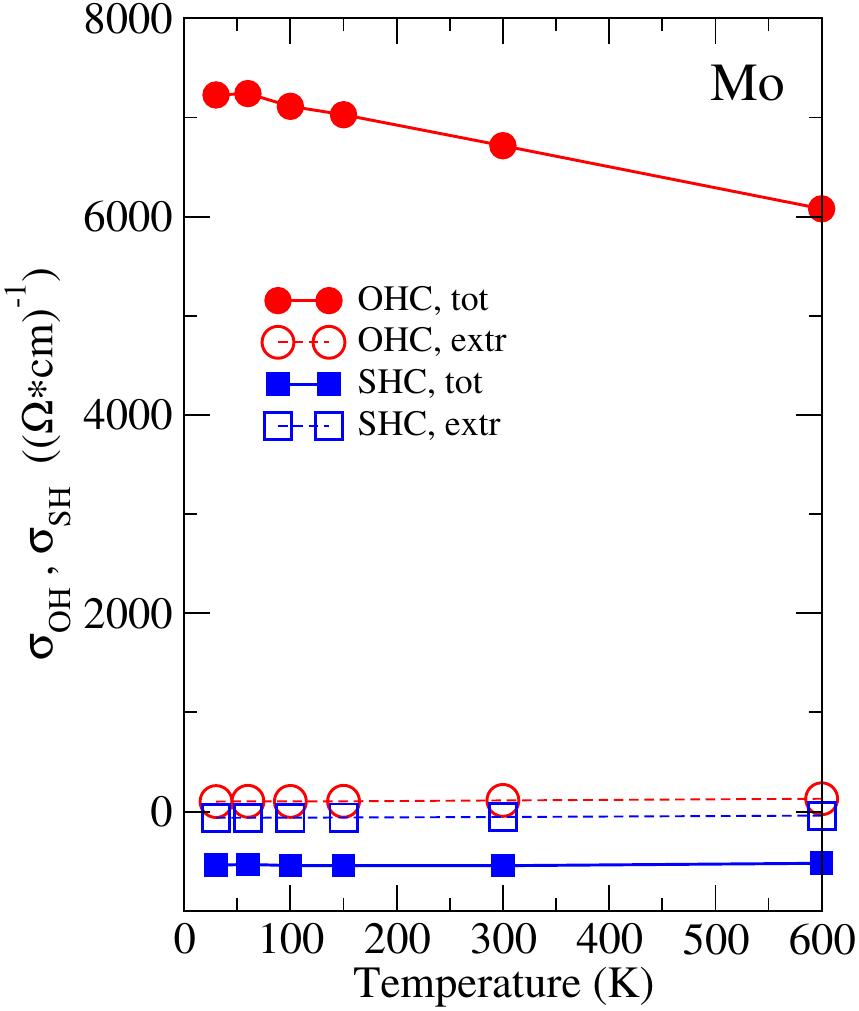}
 \includegraphics[angle=0,width=0.22\linewidth,clip]{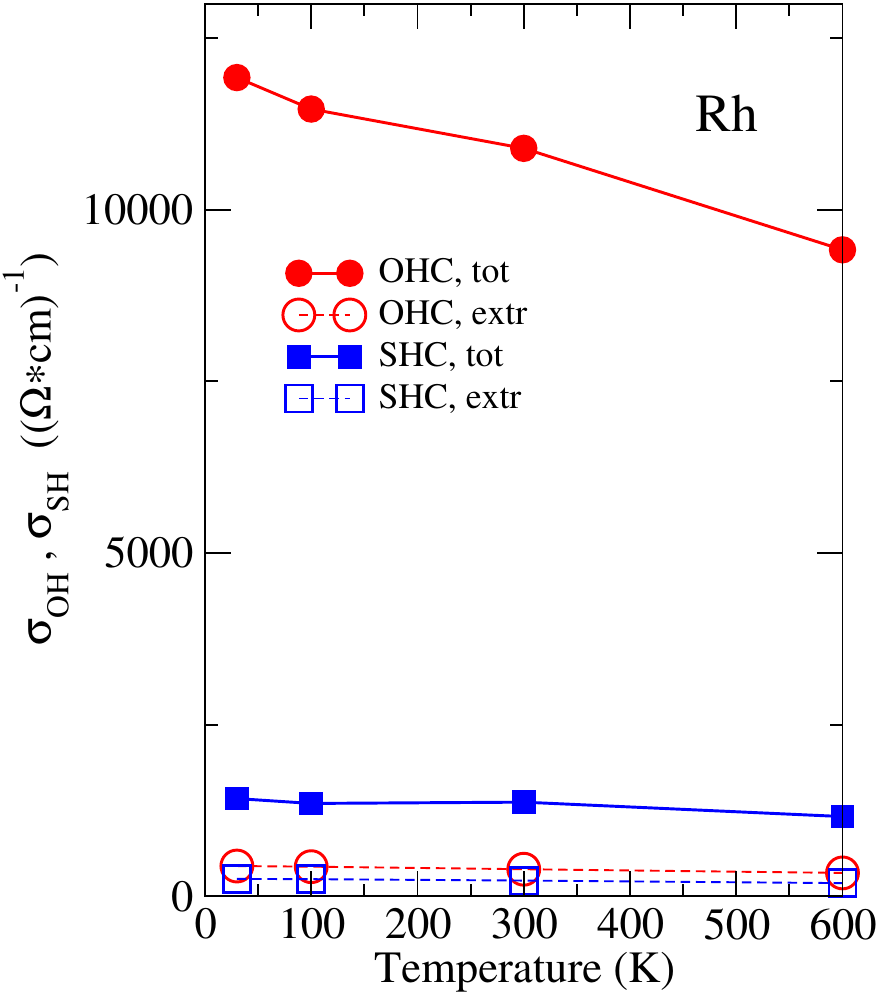} 
 \includegraphics[angle=0,width=0.22\linewidth,clip]{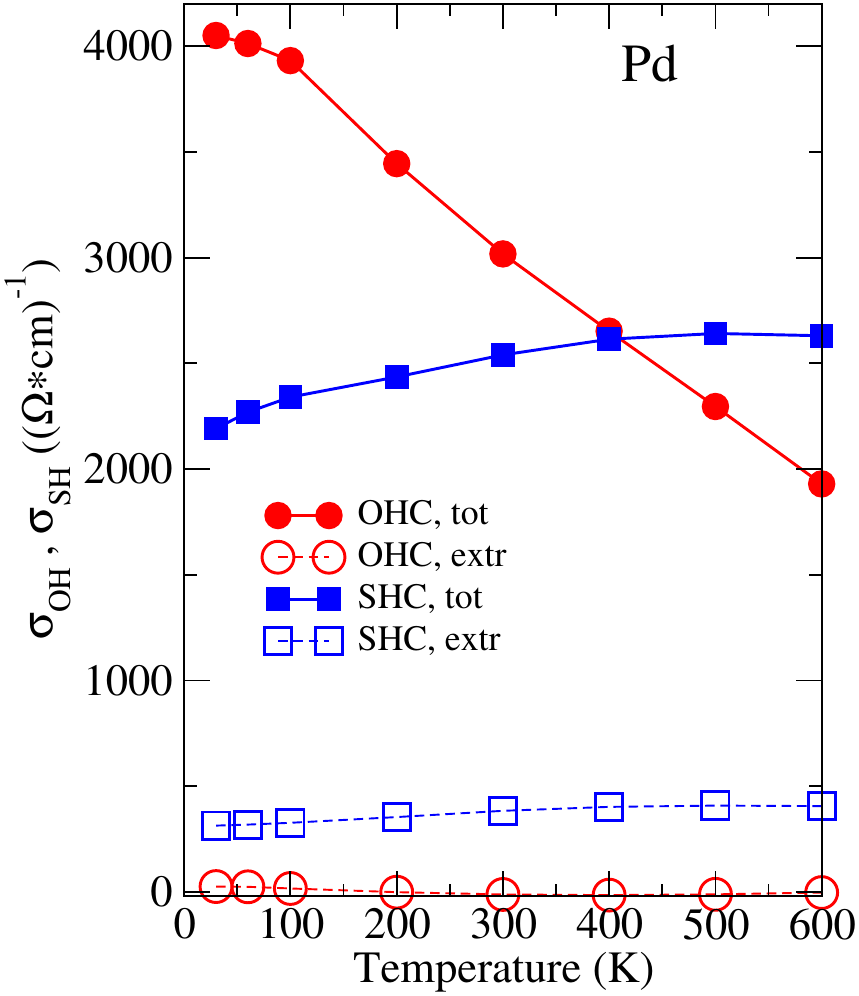}
 \includegraphics[angle=0,width=0.22\linewidth,clip]{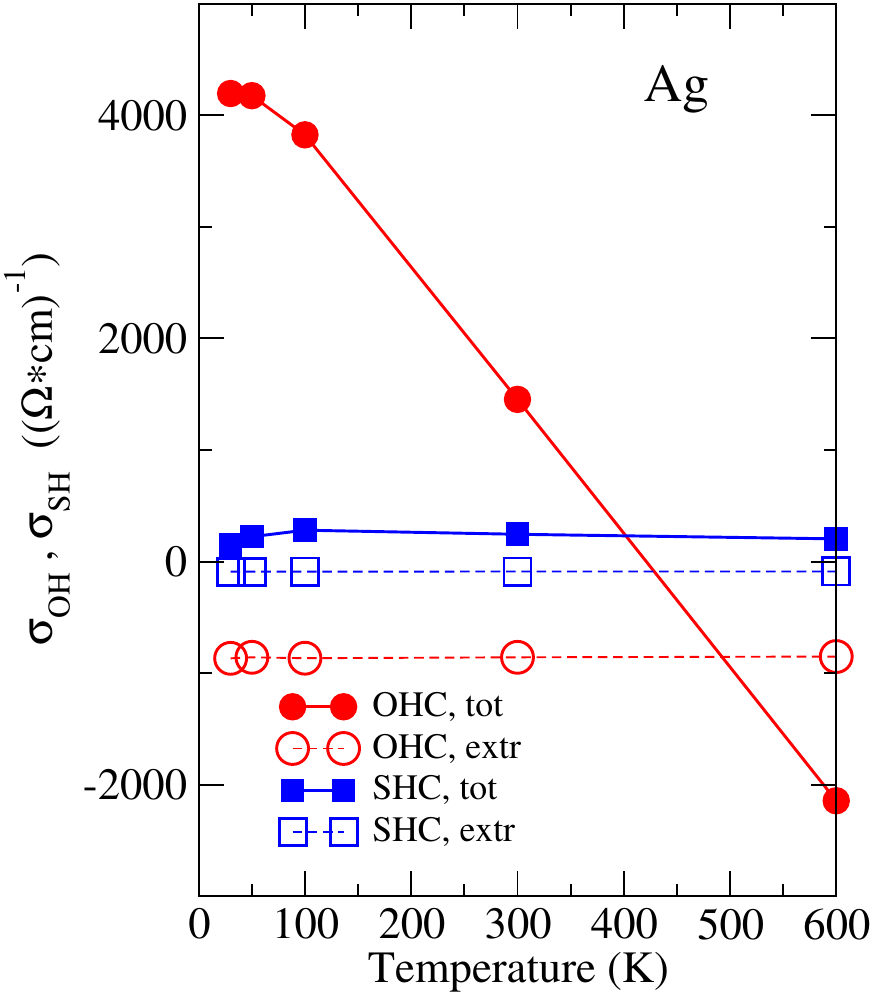}
 \includegraphics[angle=0,width=0.22\linewidth,clip]{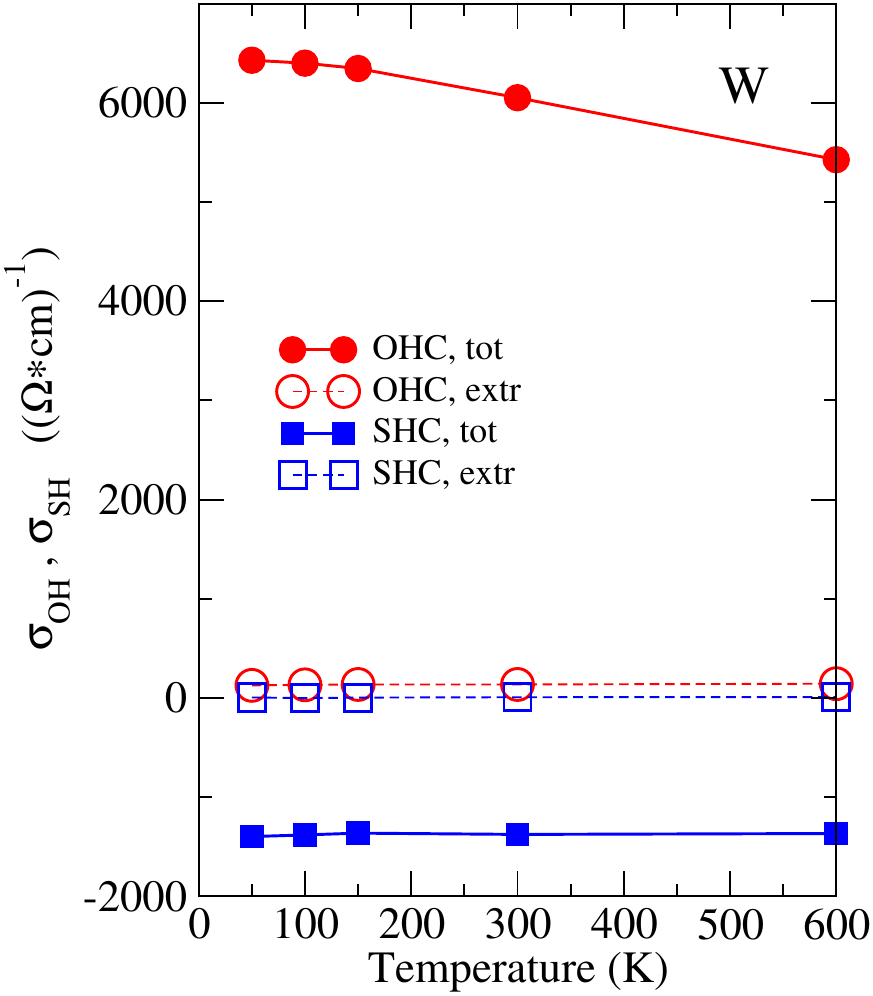}
 \includegraphics[angle=0,width=0.22\linewidth,clip]{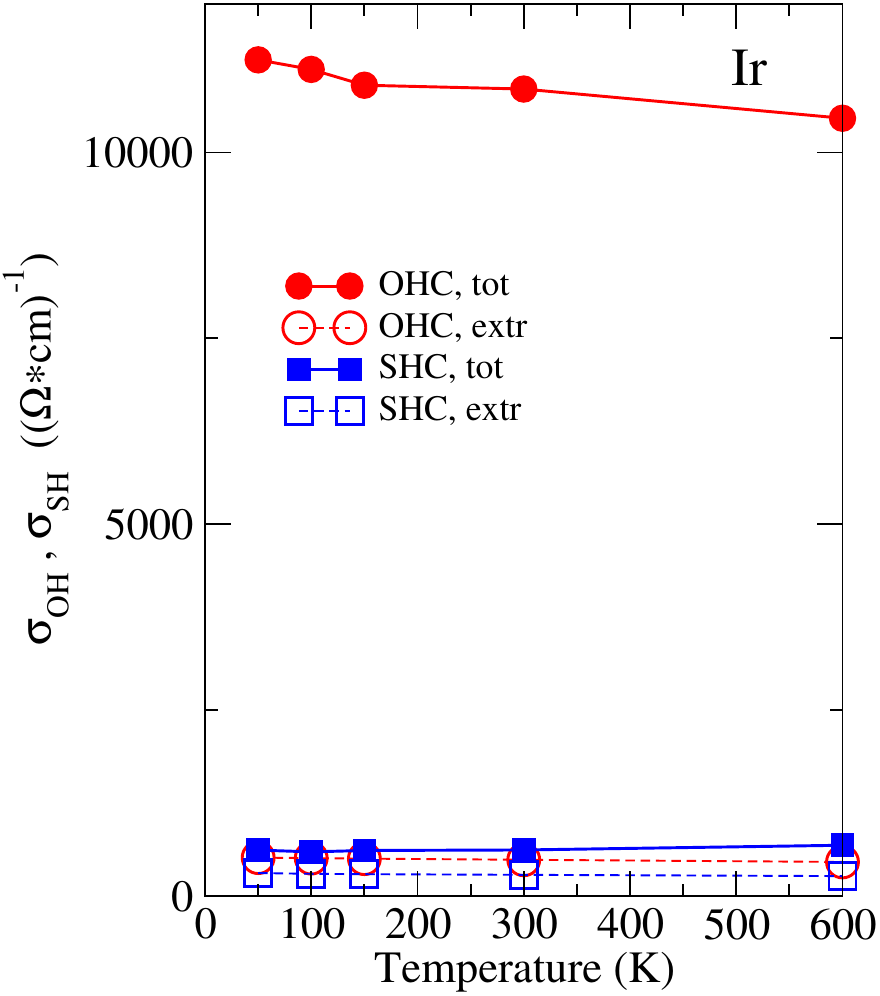}
 \includegraphics[angle=0,width=0.22\linewidth,clip]{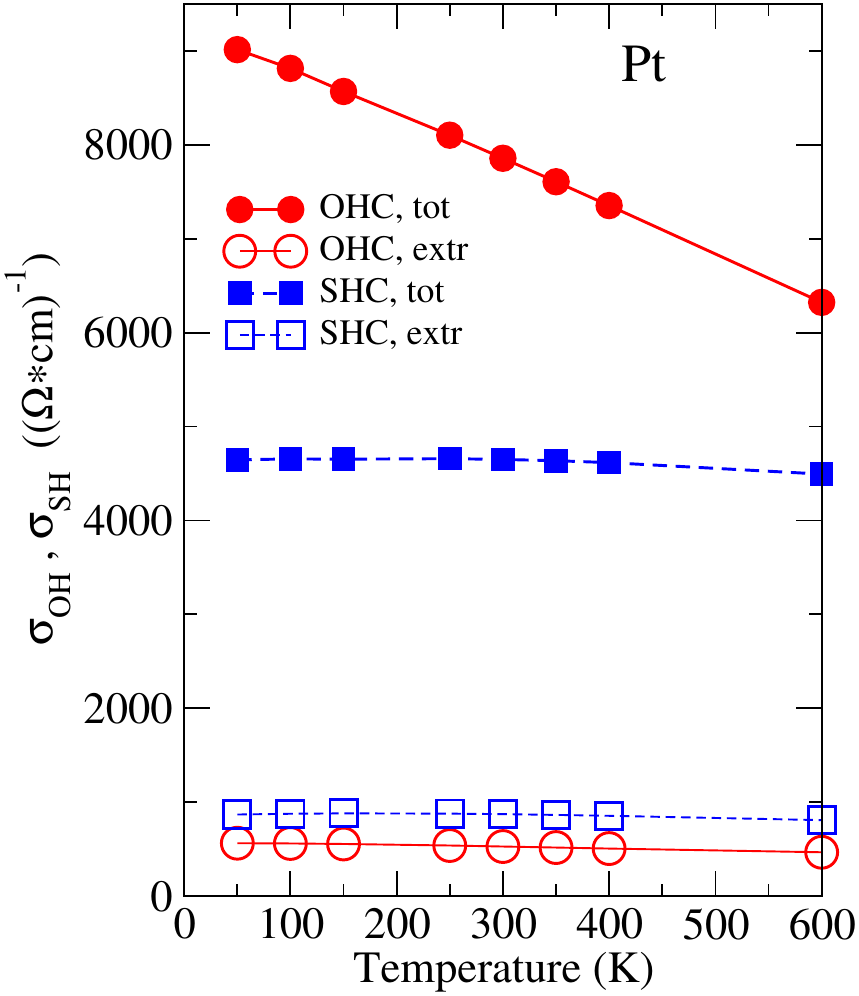}
 \includegraphics[angle=0,width=0.22\linewidth,clip]{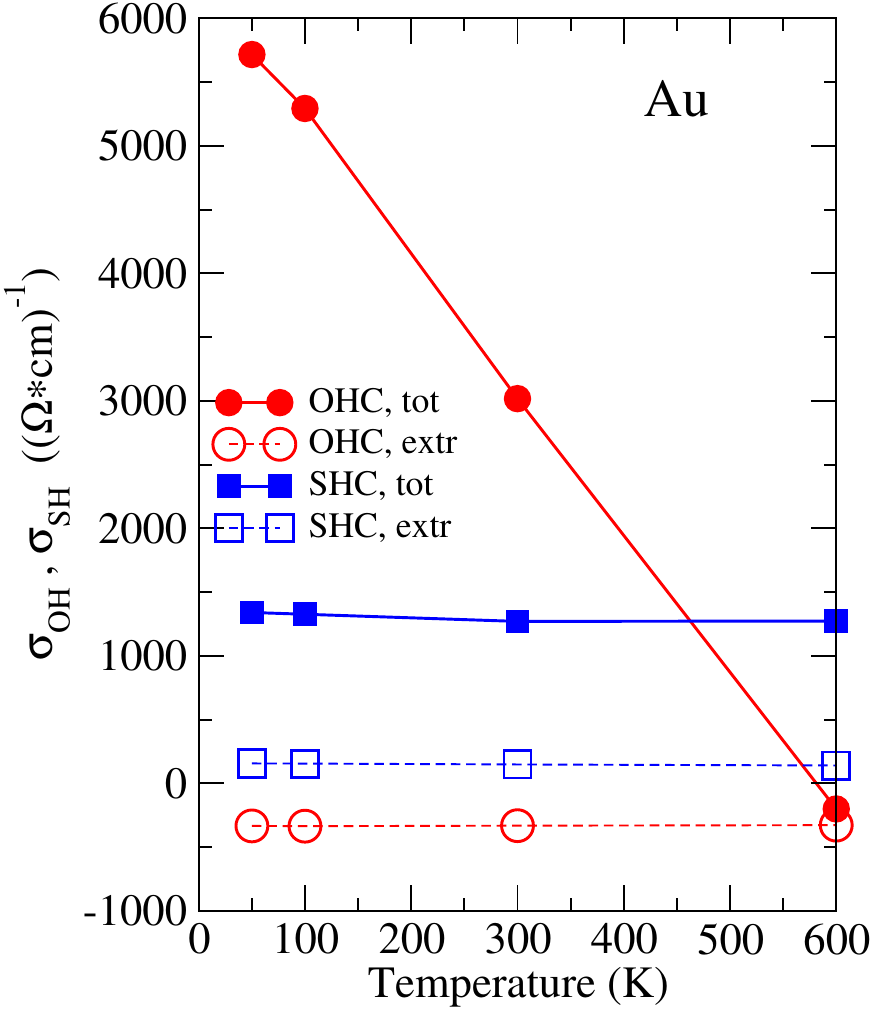}
  \caption{\label{Pure_OHE-vs-T} OH  and SH conductivities
    (full red circles and full blue squares, respectively) for selected 
    elemental $3d$, $4d$ and $5d$ non-magnetic metals, calculated
    accounting for thermal lattice vibrations and plotted as a function
    of temperature. Empty symbols represent the  extrinsic contributions
    to the OHC  and SHC.  
 }
\end{center}
\end{figure*}

The effect of a finite temperature on the OHC and SHC of pure transition metals
has been investigated in some details for temperatures up
to 600 K. 
Fig.\ \ref{Pure_OHE-vs-T} shows the total (full circles) and extrinsic
(open circles) OHC for some selected transition metals, plotted as a function of
temperature. Note that the latter quantity is determined by the Fermi
surface contribution
${\sigma}_{OH}^{{\rm extr}} = {\sigma}^{{\rm VC}}_{\rm 1,OH}(E_F) -
{\sigma}^{{\rm NVC}}_{\rm 1,OH}(E_F)$, while vertex corrections to the 
Fermi sea contribution to the OHE are negligibly small, in line with
the findings in Ref.\ [\onlinecite{TKD19}] (see also the comments in the Appendix).   
Similar plots are done also for the total (full squares) and extrinsic
(open squares) SHC.

\begin{figure}
 \begin{center}
 \includegraphics[angle=0,width=0.44\linewidth,clip]{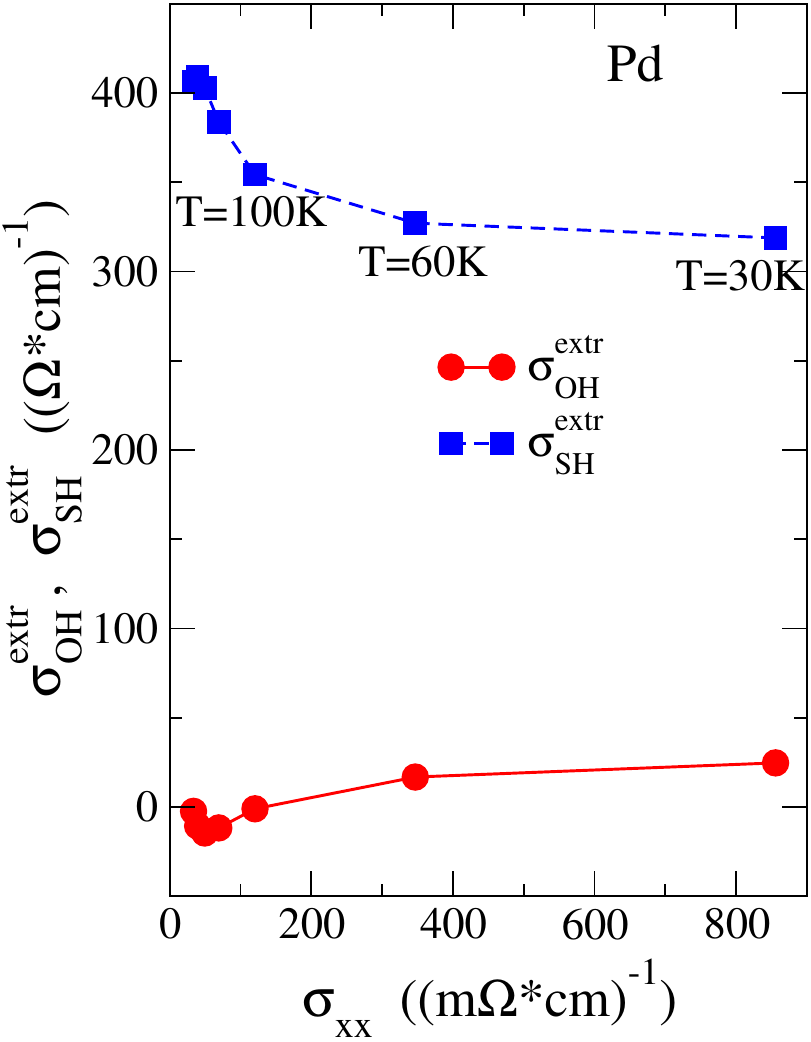}
 \includegraphics[angle=0,width=0.45\linewidth,clip]{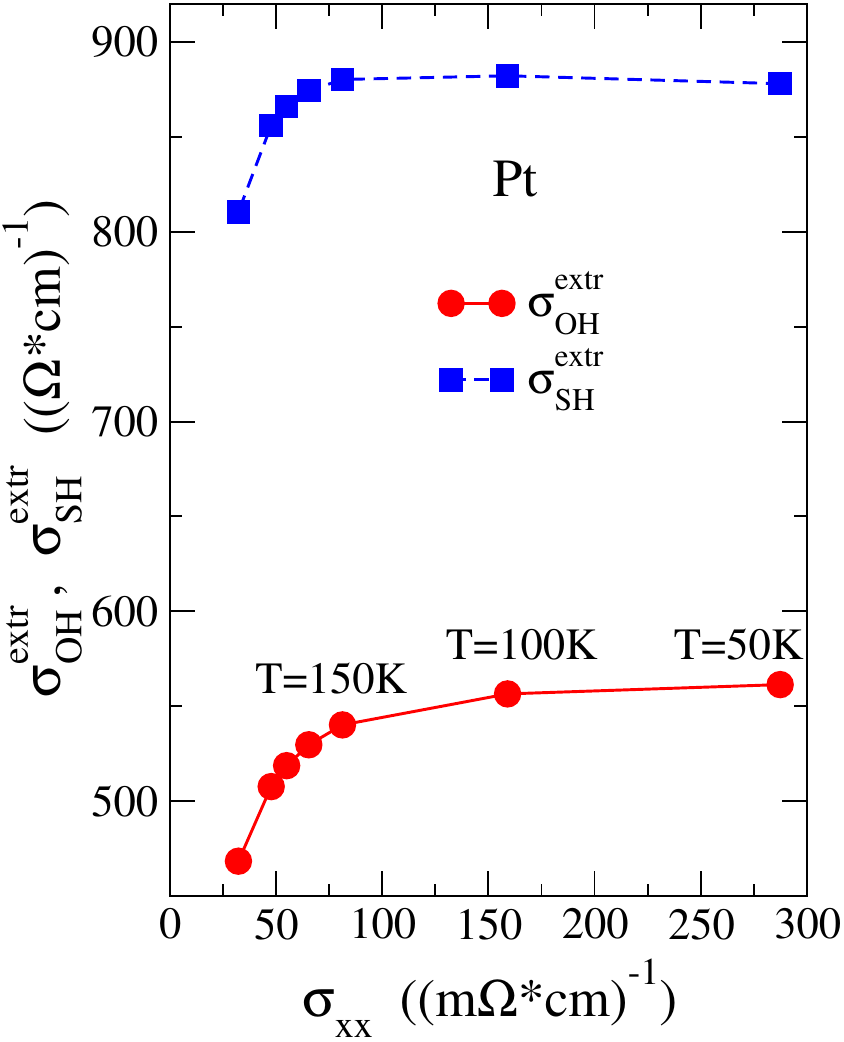}
 \caption{\label{OHC-SHC-sj-vs-T} Temperature dependent extrinsic OH and
   SH conductivities ${\sigma}^{{\rm extr}}_{\rm OH}(T)$ and
   ${\sigma}^{{\rm extr}}_{\rm SH}(T)$, for Pd and Pt, represented as a
   function of electrical conductivity.  }
\end{center}
\end{figure}

  For all systems the intrinsic OHC and SHC are dominating in the
  low-temperature limit. The extrinsic contributions are almost
  unchanged for the temperature window shown in Fig.\ \ref{Pure_OHE-vs-T}.
  This should imply a negligibly small impact of the electron-phonon skew
  scattering events on the extrinsic OHC and SHC. Otherwise, skew
  scattering would lead to their diverging behavior in the limit of $T
  = 0$ K similar to the electrical conductivity.
  Accordingly, one can conclude that the extrinsic OHC and SHC
  can be attributed primarily to the impact of the side jump scattering events.
  Interestingly, this contribution practically does not depend on
  temperature, similar to weak concentration dependence of 
  the side jump SHC in doped systems. However, one has to keep in mind
  that such an analogy is not one-to-one.
  In Fig.\ \ref{OHC-SHC-sj-vs-T} the extrinsic OHC, $\sigma_{\rm
    OH}^{\rm extr}$, and SHC, $\sigma_{\rm SH}^{\rm extr}$, are plotted for Pd and
  Pt versus the electrical conductivity $\sigma_{\mu\mu}$.
  Due to the relationship
\begin{equation} \label{Eq_4}
  \sigma_{\rm OH/SH}^{\rm extr} = \sigma_{\rm OH/SH}^{\rm skew} + \sigma_{\rm OH/SH}^{\rm sj } = 
   S_{\rm OH/SH}\sigma_{\mu\mu} + \sigma_{\rm OH/SH}^{\rm sj } \; , 
\end{equation}
  an extrapolation to $\sigma_{\mu\mu} = 0$ gives access to the
  electron-phonon side-jump contributions to the OHC and SHC in
  the low-temperature limit. This is a dominating contribution to the 
  finite-temperature extrinsic OHC and SHC in Pt and only to the SHC in
  Pd while both the side-jump and skew scattering parts are extremely
  small. Note that the side-jump contributions strongly depend on
  the material, as it follows from the results shown in
  Fig.\ \ref{Pure_OHE-vs-T}.  
  This brings us to a conclusion concerning the properties of the 
  OHC and SHC originating from thermal lattice vibrations: 
  a contribution due to skew scattering to the extrinsic OHC and
  SHC in pure metals is rather small, and the extrinsic OHC and SHC are
  dominated by the side-jump scattering mechanism. 
 These results are in line with theoretical predictions for the
 anomalous Hall effect, reported  
  by Cr\'epieux and Bruno \cite{CB01}. They attributed this effect to the
  fluctuating sign of the scattering potential associated with thermal
  phonons, which leads to a canceling effect of skew scattering
  changing sign together with the scattering potential \cite{BB80}.
  
  Note that only few experimental results have been reported for the
  temperature dependence of the OHC for pure metals.
 While a dominating intrinsic contribution to the SHC has been
 found for Pt \cite{IVH+15}, in line with the results of our calculations,
 the temperature dependent variation of the OHC in Au \cite{IVH+15} as
 well as in Pt \cite{KGL+18} was associated with the electron-phonon skew
scattering. 
It is worth noting, however, that the samples used in these experiments are
characterized by a finite residual resistivity. 
This is a crucial point,
that will be discussed below, since the impact of electron-phonon scattering
on the temperature dependence of the OHC and SHC strongly changes in the
case of doped systems.
Of course, one cannot exclude
some discrepancies between theory and experiment, originating from the
approximations used in the calculations.

\begin{figure}
 \includegraphics[angle=0,width=0.9\linewidth,clip]{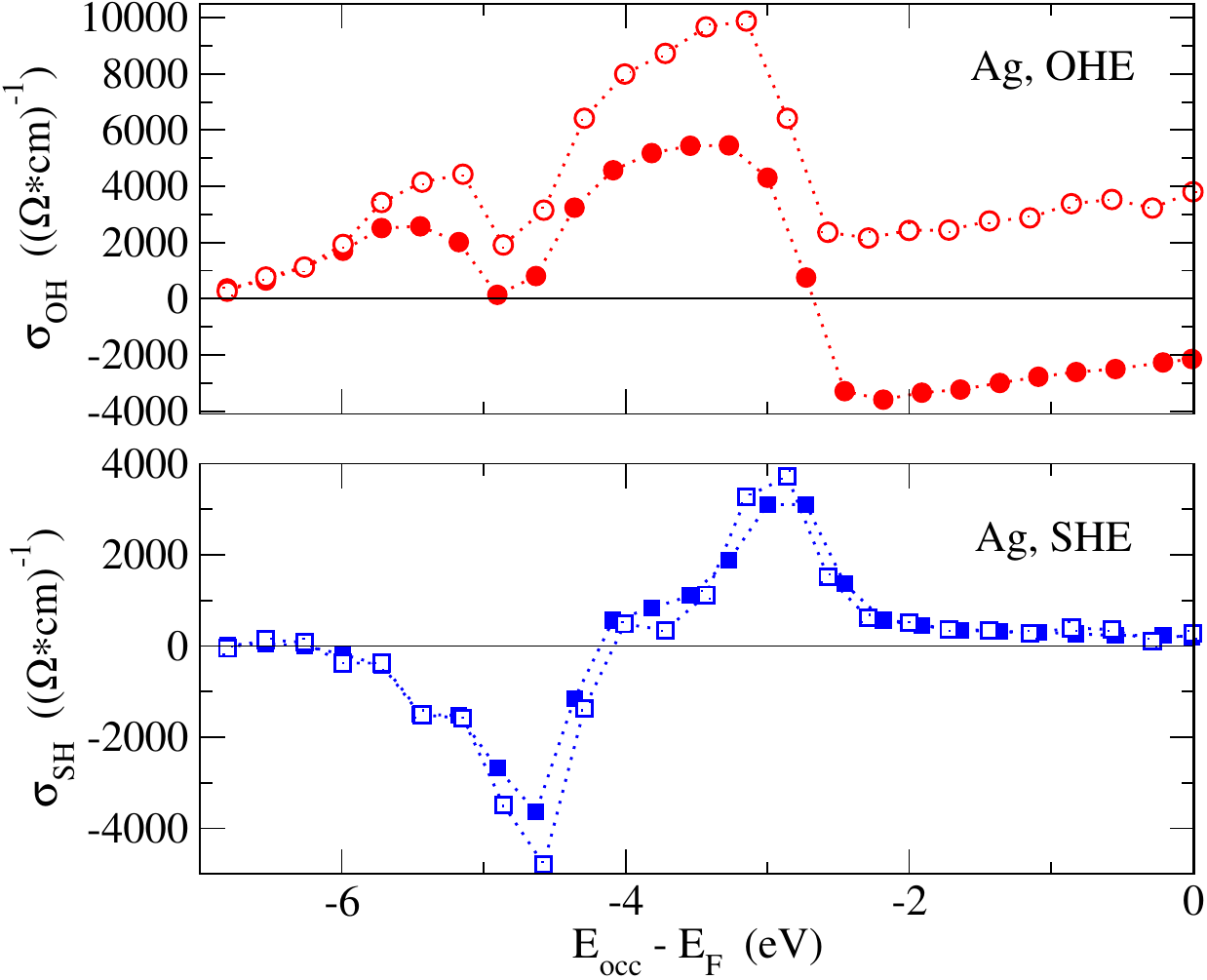}\,(a)
 \includegraphics[angle=270,width=0.45\linewidth,clip]{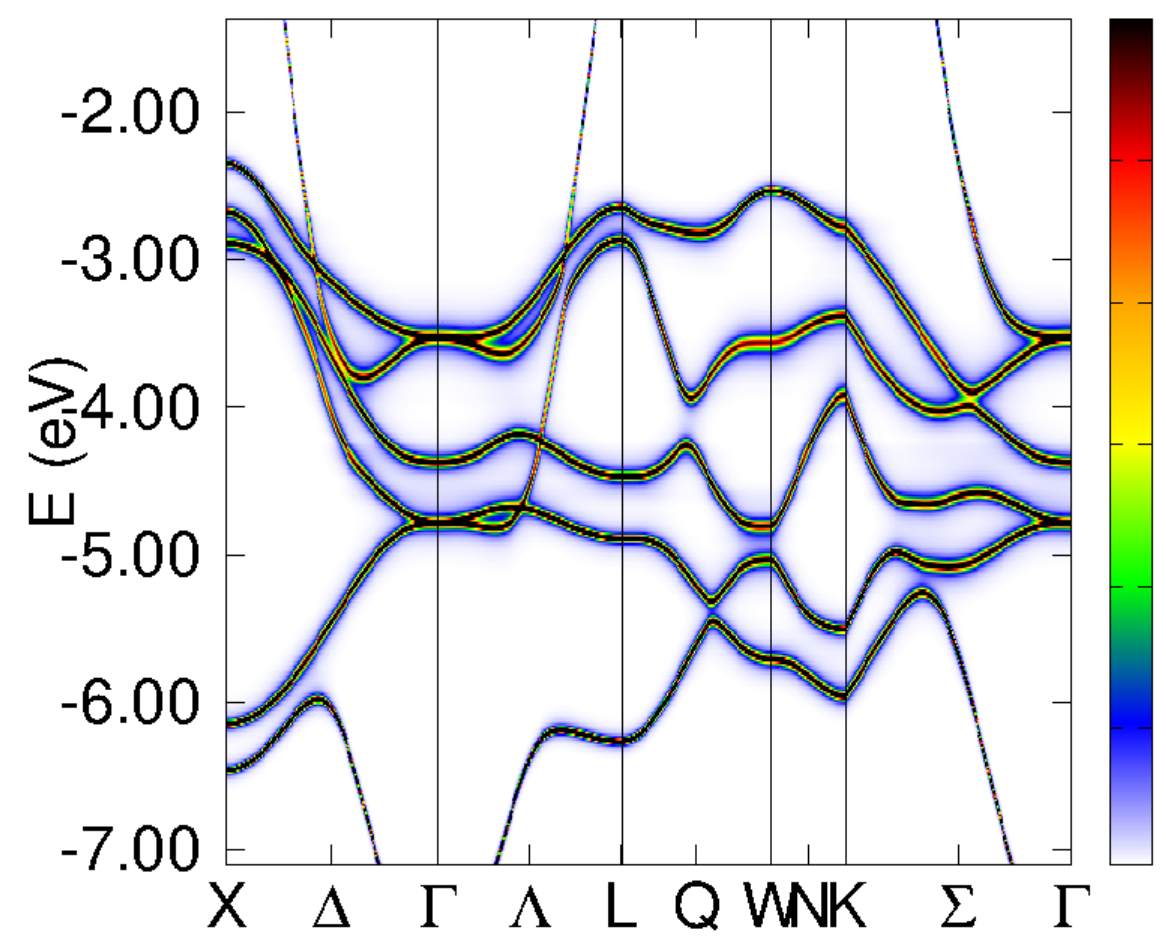}
 \includegraphics[angle=270,width=0.45\linewidth,clip]{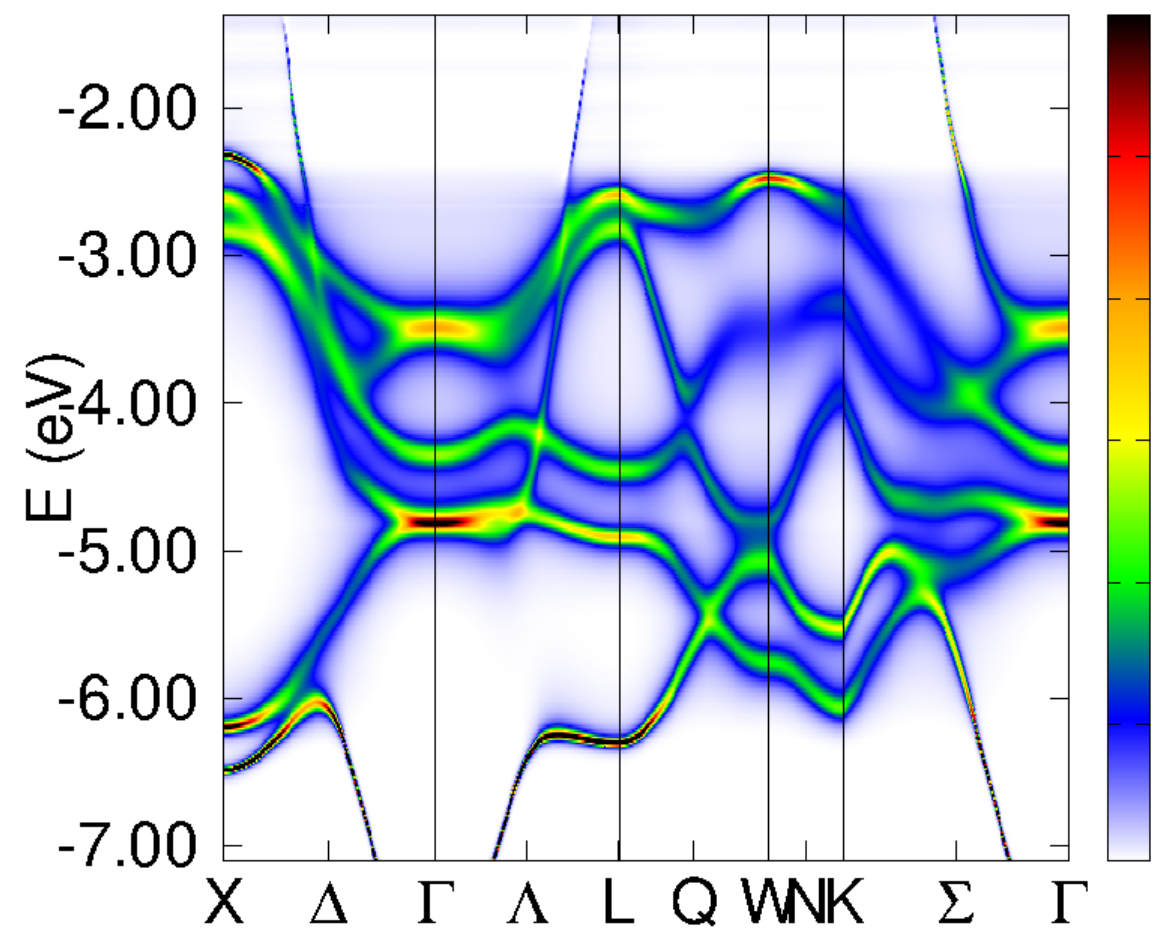} \\
\;\;\;  (b) \;\;\;\;\;\; \;\;\;\;\;\;\;\; \;\;\;\;\;\;\;\; \;\;\;\;\; \; \;   (c)
 \caption{\label{OHC-SHC-vs-EF} (a) The OHC, ${\sigma}^{{}}_{\rm OH}(E_{\rm occ})$
   (top) and SHC, ${\sigma}^{{}}_{\rm SH}(E_{\rm occ})$ (bottom) 
   calculated for Ag at $T = 100$ (open symbols) and $T = 600$ K (full
   symbols), represented as a function of occupation of electron states
   $E_{\rm occ}$. (b) and (c) The BSF for Ag at $T = 100$ and $T = 600$
   K, respectively.    
  }
\end{figure}

      Thus, the obtained results show essential 
      temperature-dependent changes of the intrinsic OHC in pure
      transition metals, associated with corresponding changes
      of the Fermi sea contribution, while the Fermi surface part
      exhibits only a weak variation with temperature. This can be
      attributed to the changes of the electronic structure arising due
      to thermal lattice vibrations leading first of all to a smearing of
      the energy bands implying a decreasing lifetime of the electronic
      states.  
      In contrast to the  OHC, a weak temperature-induced change can be
      seen for both the intrinsic and extrinsic contributions to the SHC.
      To gain insight into to origin for this different behavior 
      we have performed calculations for the OHC and SHC as a
      function of the occupation of the electron states, using Ag as a
      representative system as it has a rather strong variations of
      the OHC with temperature.
      In the calculations, the occupation is controlled by the upper energy
      limit $E_{\rm occ}$ which is equal to $E_F$ under normal
      condition. The calculations have been done for two lattice temperatures,
      $T = 100$ and $T = 600$~K. Fig. \ref{OHC-SHC-vs-EF}(a) shows
      the results for total OHC and SHC, which, however, are mainly
      determined by the intrinsic contributions, as it is shown in
      Fig.\ \ref{Pure_OHE-vs-T}. One can see in
      Fig. \ref{OHC-SHC-vs-EF}(a) (bottom) a minor difference between
      the results for the SHC obtained for two different temperatures,
      that, however, is not the case for the OHC
      (Fig. \ref{OHC-SHC-vs-EF}(a), top). The difference 
      between the results for the OHC at $T = 100$ and $T = 600$~K 
      increases together with increasing occupation, i.e. $E_{\rm occ}$, of
      the $d$-states, and almost does not change when $E_{\rm occ}$ goes
      beyond the top of the $d$-band. This occupation dependence, in
      particular, may be responsible for a weaker temperature dependence
      of the OHC for the early transition metals.  
The strong temperature-induced modification of the OHC can be attributed 
to the impact of thermal lattice vibrations on the electronic structure 
leading to a strong smearing of the energy bands, increasing with 
temperature. This can be seen in the bottom panel of
Fig.\ \ref{OHC-SHC-vs-EF} that gives the
Bloch spectral function $A(\vec{k},E,T)$ for Ag, calculated for $T =
100$ (b)
and $T = 600$\,K (c). Furthermore, it is crucial that the
temperature-induced lattice distortion breaks local symmetry at each lattice site.
As soon as the origin of the OHE is associated with the
$\vec{k}$-dependent orbital texture controlled by symmetry
\cite{KTH+09,GJL+21,GLO+24,TB24,HLK23}, the OHC variation at finite
temperature may stem from broken local symmetry at every atomic
position, leading to a modification of the orbital texture, increasing
with temperature (see also discussions in Ref.\ [\onlinecite{TB24}]).  
Interestingly, only minor changes occur for the SHC, in conflict with
the idea that the SOC-driven spin Hall current density 
originates from the orbital Hall current density\cite{KTH+09}.
This indicates in particular that the relationship between these
quantities is not straightforward.

\subsection{Doped materials}

\subsubsection{Doping effect in OHC and SHC.}

Next, we discuss the properties of the OHC in the presence of disorder
in doped materials, and compare the results with 
corresponding properties of the SHC.

\begin{figure}
 \begin{center}
 \includegraphics[angle=0,width=0.43\linewidth,clip]{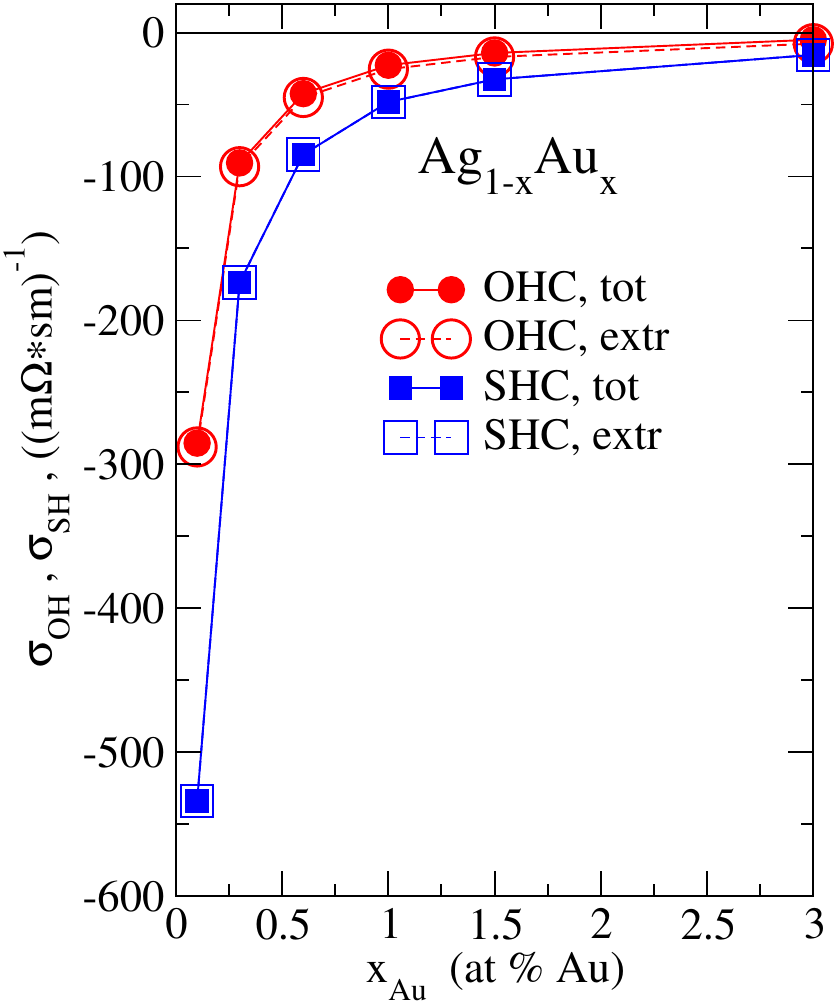}\,(a)
 \includegraphics[angle=0,width=0.40\linewidth,clip]{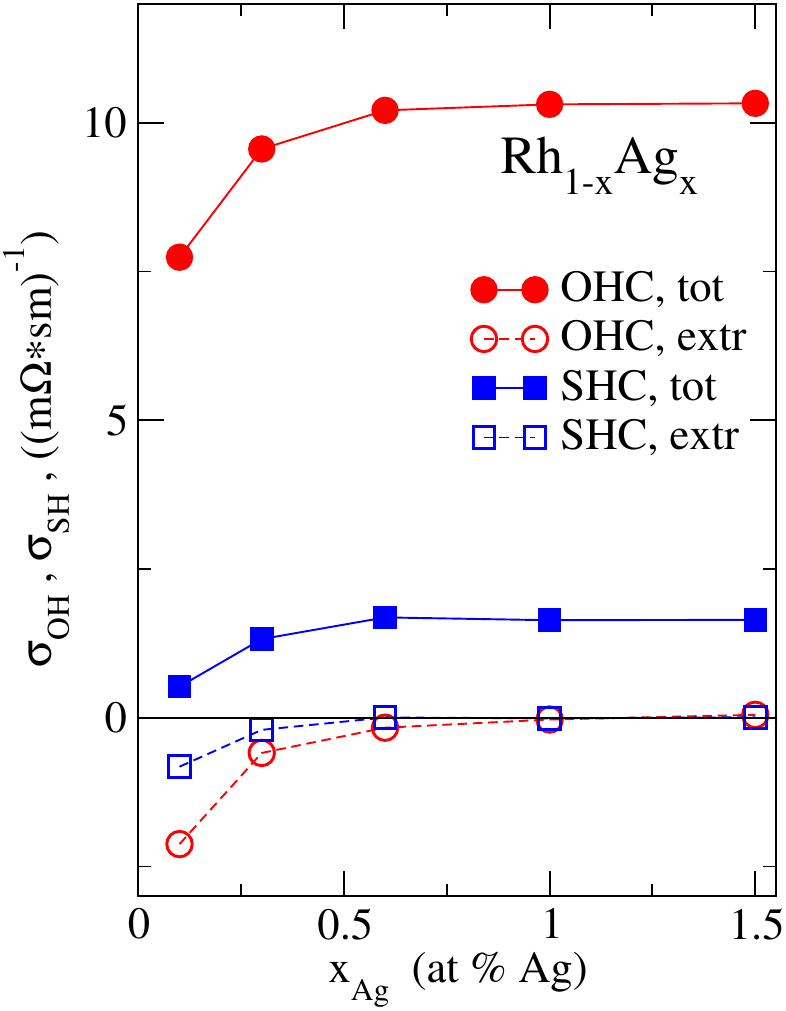}\,(b)
 \includegraphics[angle=0,width=0.43\linewidth,clip]{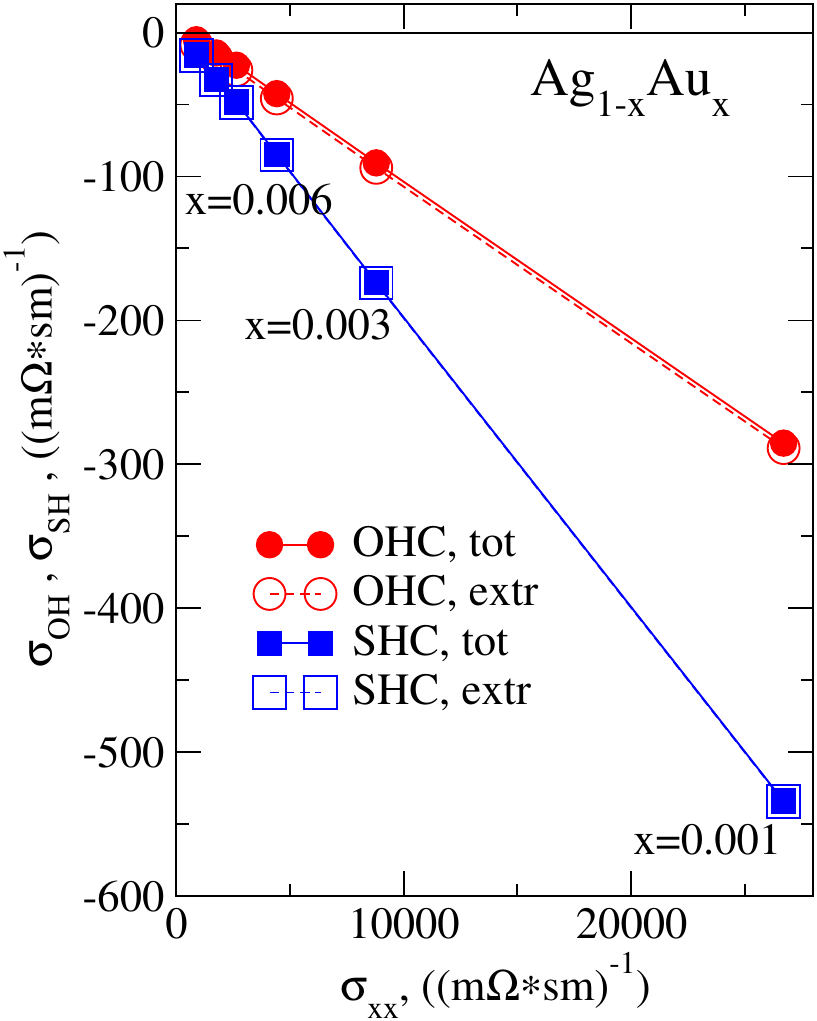}\,(c)
 \includegraphics[angle=0,width=0.40\linewidth,clip]{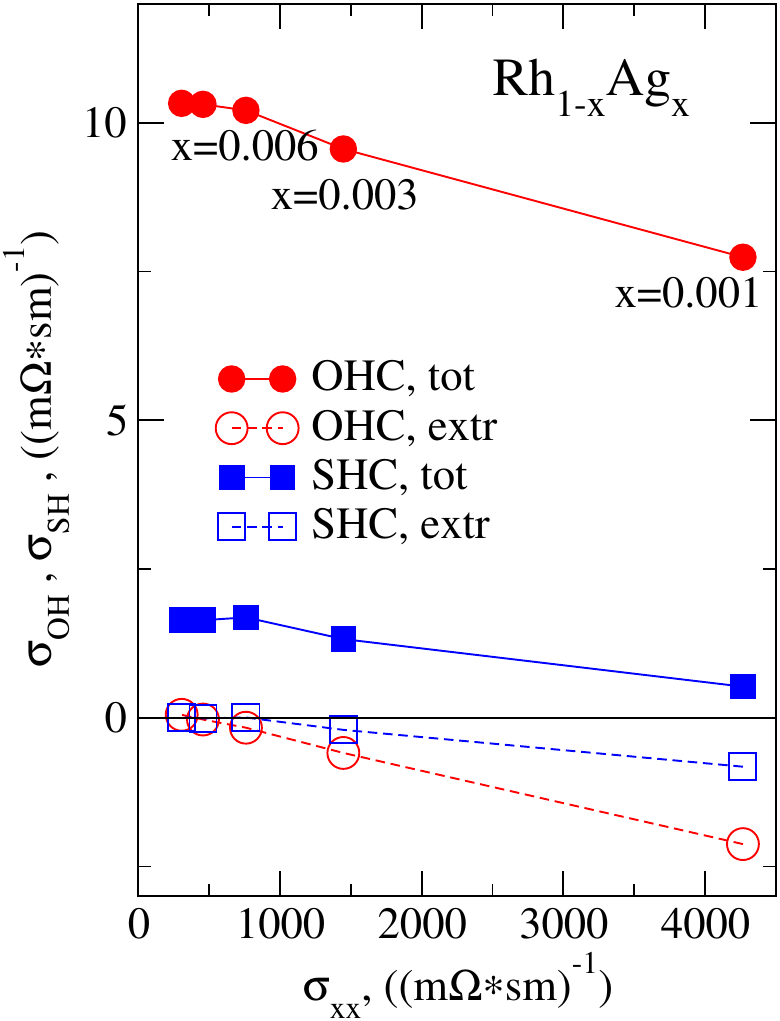}\,(d)
  \caption{\label{A-B_x-OHE_SHE-vs-x}  OH (circles) and SH (squares)
    conductivities for  Ag$_{1-x}$Au$_{x}$ (a) and Rh$_{1-x}$Ag$_{x}$ (b) around
    the Ag-rich and Rh-rich limits, respectively, as a function of the
    impurity concentration. OH (circles) and SH
    (squares) conductivities plotted as a function of the electrical
    conductivity for Ag$_{1-x}$Au$_{x}$ (c)  and  Rh$_{1-x}$Ag$_{x}$
    (d), respectively. Full symbols in all figures correspond to the
    total OHC and SHC, while 
    open symbols represent the extrinsic contributions. } 
\end{center}
\end{figure}

We consider first, two arbitrary chosen doped systems: (i) one with the
host and impurity atoms corresponding to different transition metal periods
but having the same number of valence 
electrons, Ag$_{1-x}$Au$_{x}$, and (ii) another one with the host and
impurity atoms which belong to the same periods but with different
number of valence electron, Rh$_{1-x}$Ag$_{x}$. The corresponding
results for the OHC and SHC are plotted in
Fig.\ \ref{A-B_x-OHE_SHE-vs-x}(a) and (b) (full circles for the OHC and
full squares for the SHC), respectively, as a function of
concentration. One can see a divergent behavior for the OHC and SHC for
Ag$_{1-x}$Au$_{x}$ towards the pure 
limit, which however is hardly seen in the case of Rh$_{1-x}$Ag$_{x}$. The
reason for this difference is discussed next.

\begin{figure}
 \begin{center}
 \includegraphics[angle=0,width=0.42\linewidth,clip]{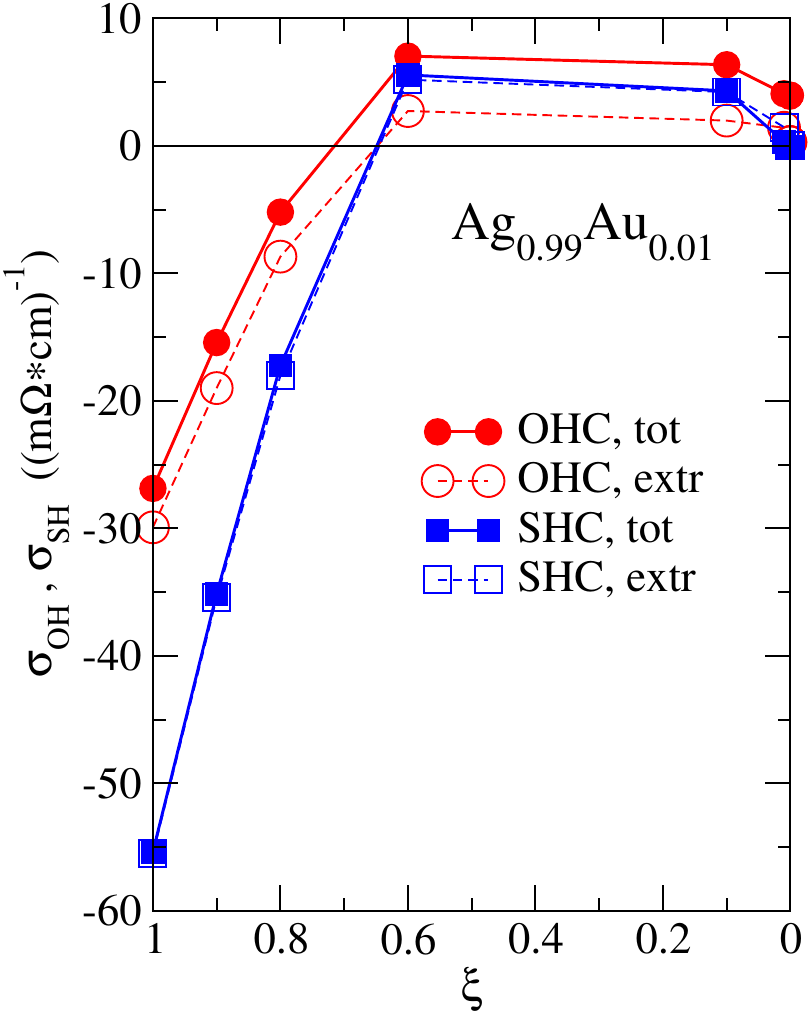}\,(a)
 \includegraphics[angle=0,width=0.42\linewidth,clip]{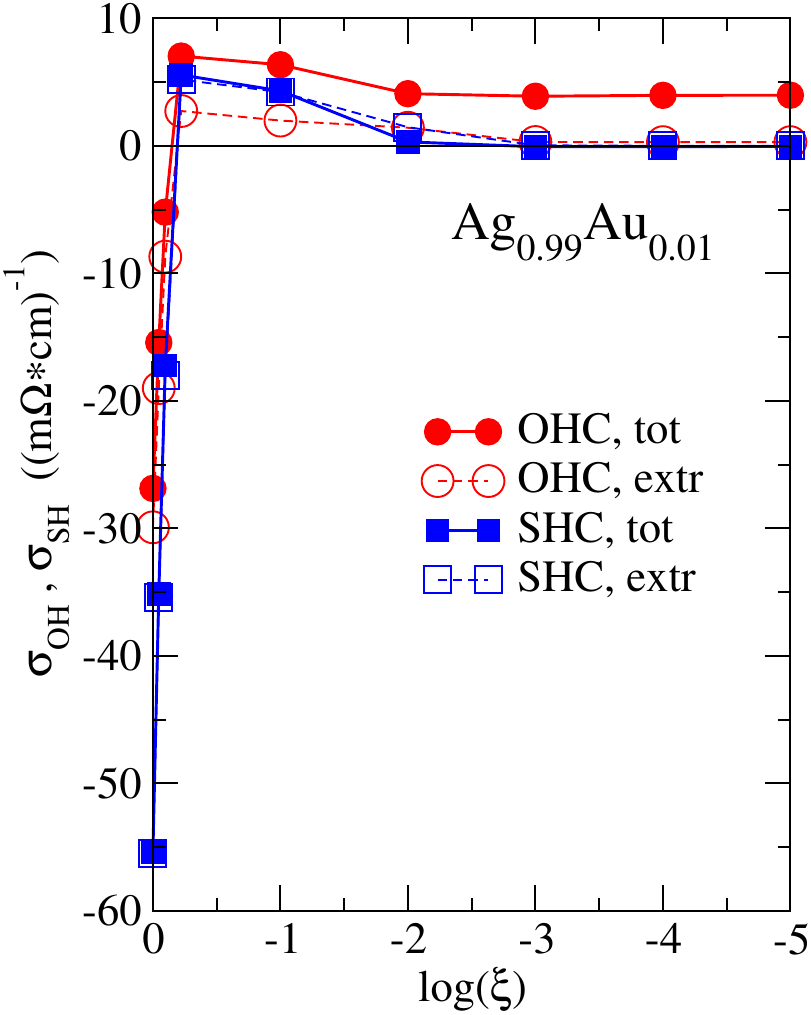}\,(b)
  \caption{\label{Ag-Au0.01_OHE_SHE-vs-SOC} The OHC and SHC
    Ag$_{0.99}$Au$_{0.01}$ as a function of the SOC scaling factor
    $\xi$. Figures (a) and (b) represent the same results  plotted
    using linear and logarithmic meshes for  $\xi$, respectively. }  
\end{center}
\end{figure}

As is shown in Fig.\ \ref{A-B_x-OHE_SHE-vs-x} (a), a strong increase of
the OHC and SHC towards the pure limit ($x \to 0$) in Ag$_{1-x}$Au$_{x}$ is mainly
determined by the extrinsic contribution arising due to 
skew-scattering. This conclusion follows from the relationship $\sigma_{\rm
  OH/SH}^{\rm skew} = S_{\rm OH/SH} \sigma_{\mu\mu}$ between the
skew-scattering contributions to the extrinsic OHC and SHC and the
electrical conductivity, which is demonstrated in
Fig.\  \ref{A-B_x-OHE_SHE-vs-x}(c). As one can see, the total and
extrinsic OHC and SHC are lying almost on top of each other, and the
extrapolation to the limit $\sigma_{\mu\mu} = 0$ 
(see Section \ref{Theoretical details}) leads to
small values for the intrinsic as well as side-jump extrinsic
contributions to the OHC and SHC. 
This finding gives an evidence that the same skew scattering mechanism should be
responsible for both the OHC and SHC in Ag$_{1-x}$Au$_{x}$, 
although the skewing factors are different for the OHC and SHC. 
This conclusion is also in line with the results for the SOC dependence
of the extrinsic OHC and SHC shown in
Fig.\ \ref{Ag-Au0.01_OHE_SHE-vs-SOC} for Ag$_{1-x}$Au$_{x}$, with the 
SOC strength controlled via the scaling factor $\xi$. For the sake of
convenience, the same data are plotted using a linear (a) and a 
logarithmic (b) scale for the ordinate axis.
In Fig.\ \ref{Ag-Au0.01_OHE_SHE-vs-SOC}(a) one can see a quick decrease in
magnitude for the total OHC and SHC (which are negative) followed by
a variation of
the extrinsic contributions corresponding to the SOC reduction with the
scaling factor decreasing 
from $\xi = 1$ to the value $\xi \approx 0.7$ for which both the OHC and SHC change
sign. A further decrease of the scaling factor $\xi \to 0$ leads to the
decrease of a positive extrinsic and total OHC and SHC (see
Fig.\ \ref{Ag-Au0.01_OHE_SHE-vs-SOC}(b)). In the limit of 
$\xi = 0$ (i.e. no SOC), the extrinsic OHC vanishes and the total OHC
coincides with its
intrinsic contribution. The total SHC decreases due to a decrease of 
both contributions, extrinsic and intrinsic, and vanishes in the
limit of $\xi = 0$. 

In the case of Rh$_{1-x}$Ag$_{x}$, the behavior of the OHE and SHE shown
in Fig.\ \ref{A-B_x-OHE_SHE-vs-x} (b) (full symbols) is rather different
when compared to Ag$_{1-x}$Au$_{x}$.
First of all, the OHE and SHE at small impurity concentration are much
smaller than in the case of 
Ag$_{1-x}$Au$_{x}$. Their extrinsic contributions (open symbols) are negligibly small
for an Ag concentration increasing above  $\approx 0.5$
at.\%, where the OHE and SHE are mainly determined by the intrinsic
contributions. Furthermore, the extrinsic OHE and SHE are negative and
exhibit a slow increase in magnitude for $x \to 0$. This increase
correlates 
with a slow increase of the electrical conductivity $\sigma_{\mu\mu}$
and therefore can be attributed to the skew-scattering effect.   
Plotting the total $\sigma_{\rm OH}$ and $\sigma_{\rm SH}$ versus $\sigma_{\mu\mu}$ in
Fig.\ \ref{A-B_x-OHE_SHE-vs-x}(d) (full symbols) and extrapolating the
curves to the limit $\sigma_{\mu\mu} = 0$, one can easily sea the
leading role of the intrinsic contribution to the OHC and SHC in the superclean
regime. Small values for the side-jump contributions can be deduced also
from the extrapolation of the the extrinsic OHC and SHC (open symbols).

\begin{figure}
 \begin{center}
 \includegraphics[angle=0,width=0.43\linewidth,clip]{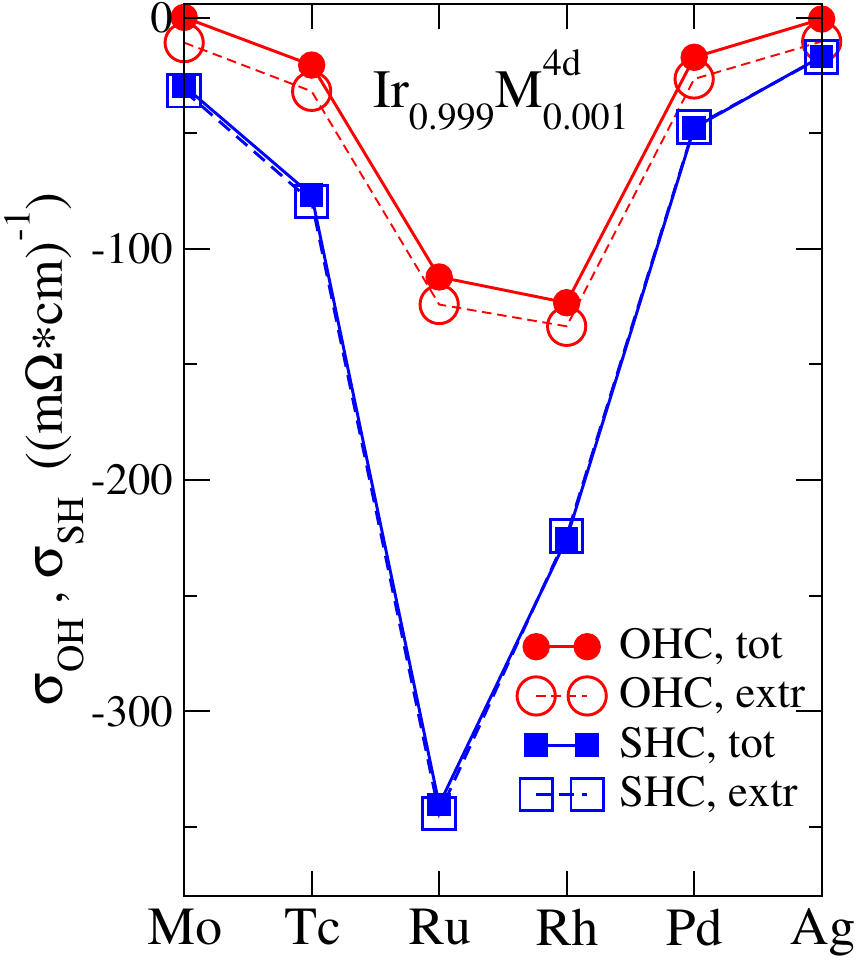}\,(a)
 \includegraphics[angle=0,width=0.43\linewidth,clip]{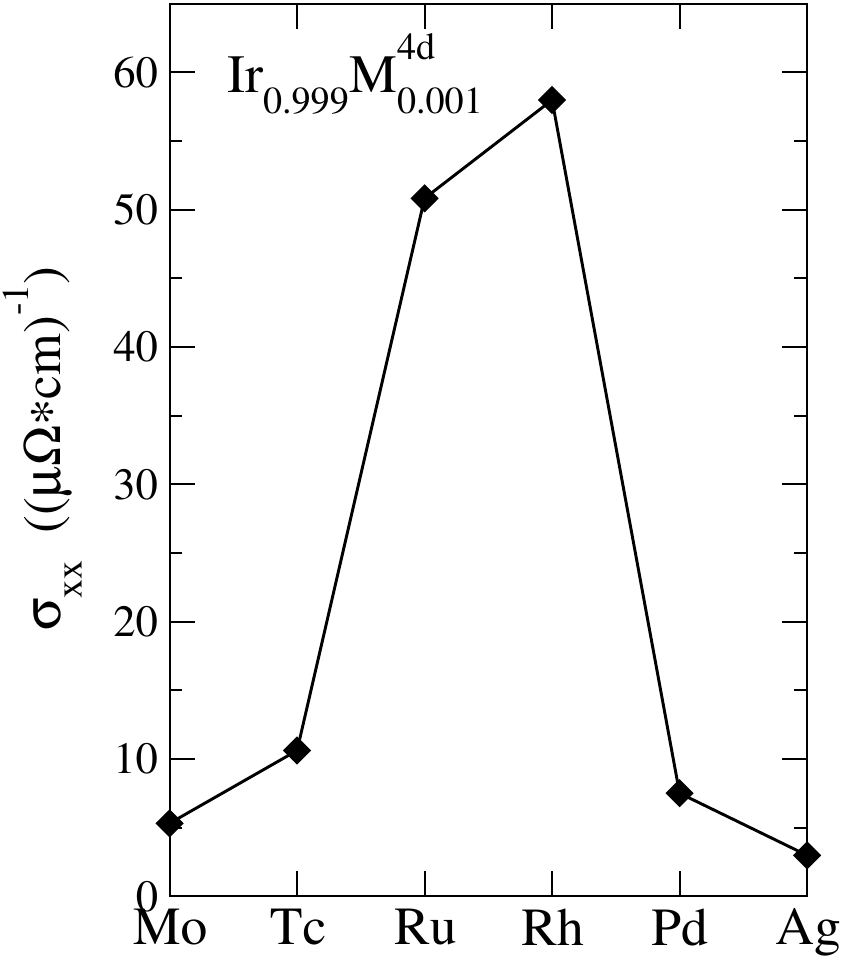}\,(b)
 \includegraphics[angle=0,width=0.43\linewidth,clip]{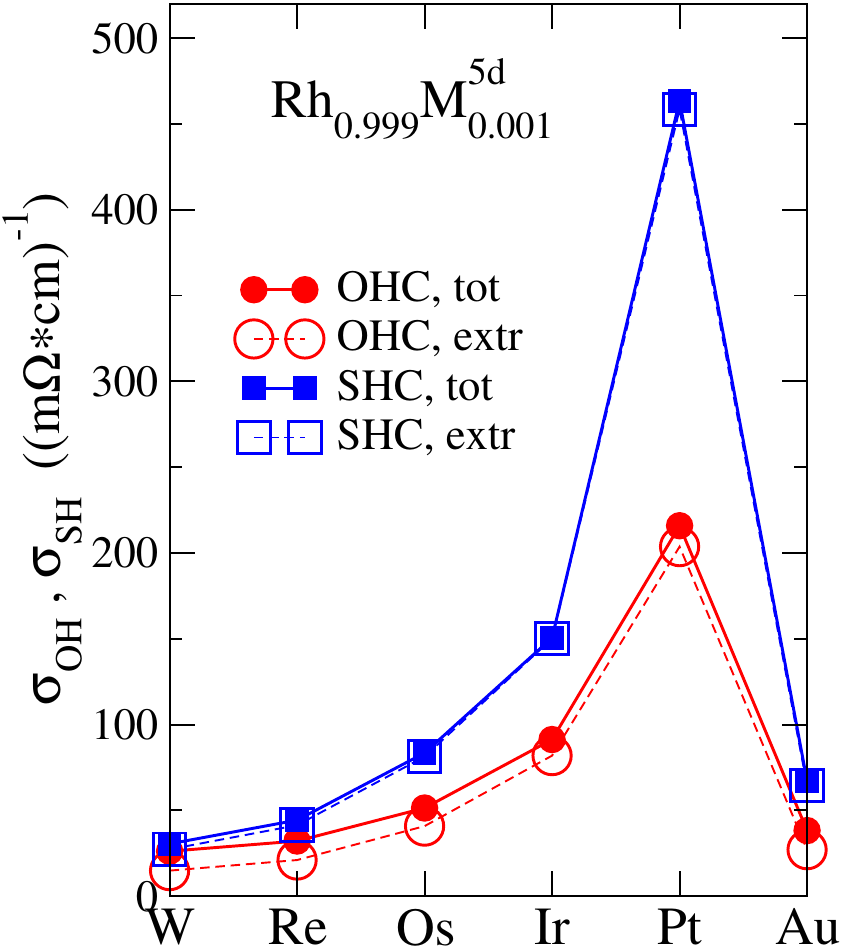}\,(c)
 \includegraphics[angle=0,width=0.43\linewidth,clip]{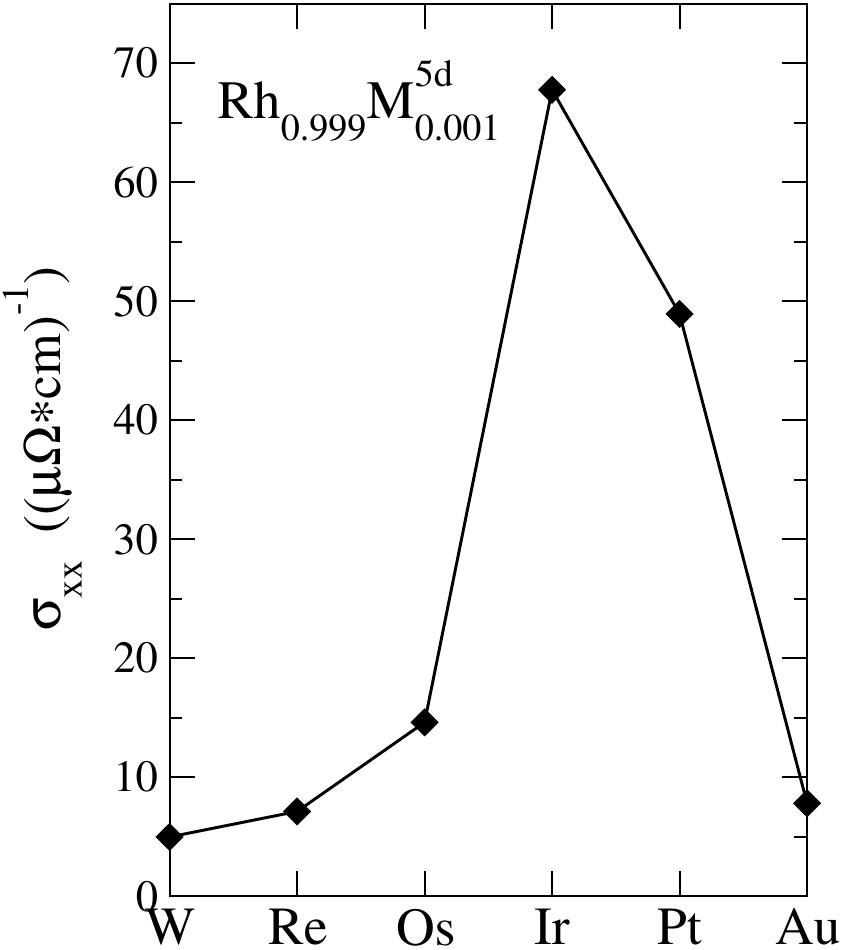}\,(d)
 \includegraphics[angle=0,width=0.43\linewidth,clip]{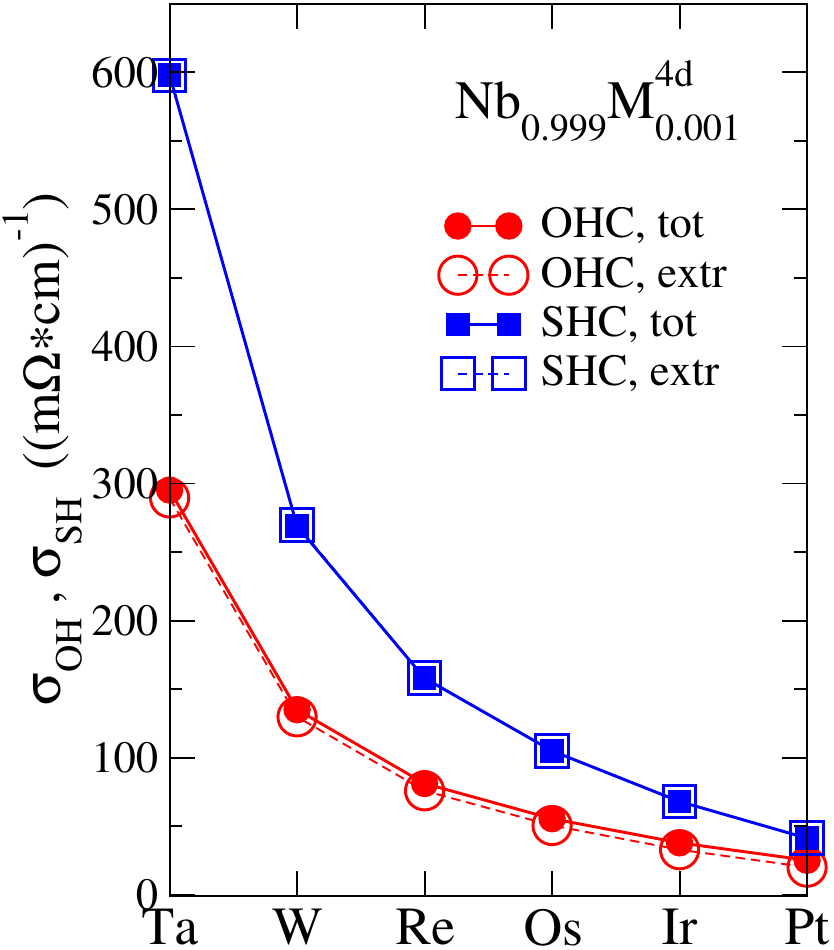}\,(e)
 \includegraphics[angle=0,width=0.43\linewidth,clip]{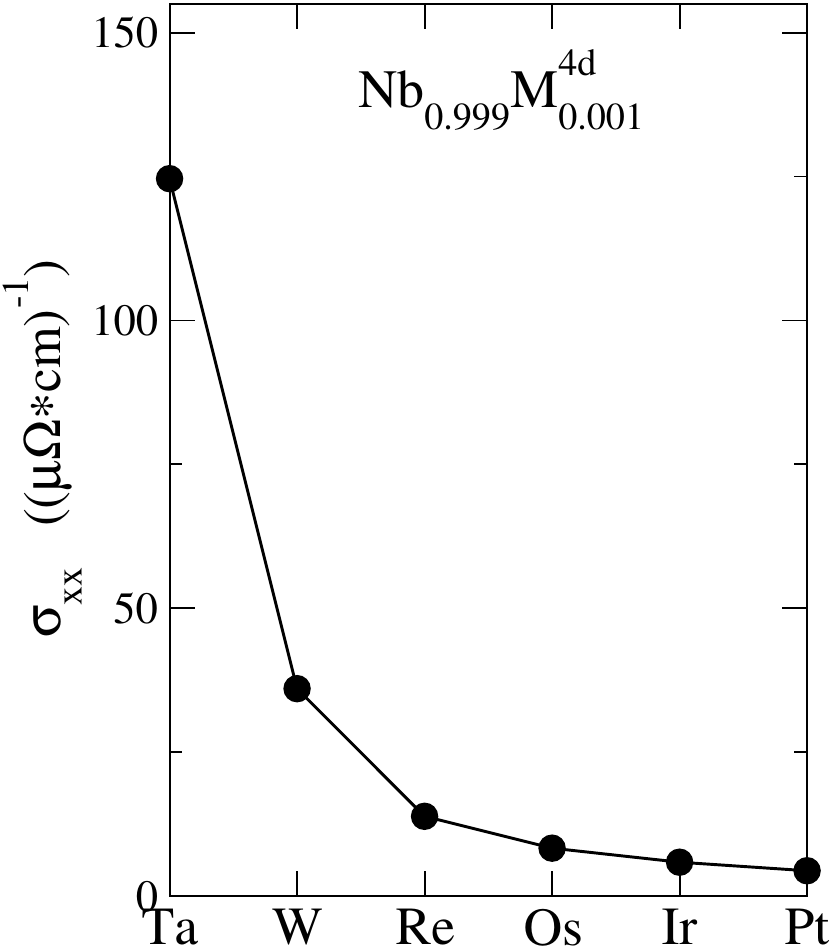}\,(f)
  \caption{\label{Me1_vs_Me2_OHE_SHE} (a), (c) and (e): OHC (circles)
    and SHC (squares) for  Ir$_{0.999}\,M^{4d}_{0.001}$, Rh$_{0.999}\,M^{5d}_{0.001}$ and
    Nb$_{0.999}\,M^{5d}_{0.001}$, where total values are shown by full
    symbols, and extrinsic contributions - by open symbols. (b), (d) and
    (f) represent corresponding results for the longitudinal electrical
    conductivity $\sigma_{\nu\nu}$.  }   
\end{center}
\end{figure}

To gain insight into the origin of such a strong difference between the
properties of the OHC and SHC of Ag$_{1-x}$Au$_{x}$ and
Rh$_{1-x}$Ag$_{x}$, we have performed a set of calculations of the OH
and SH conductivities for the materials having the same host but with
different elements used for impurities. The calculations were 
done for $T = 0$ K. We consider three representative systems,
Ir$_{0.999}\,M^{4d}_{0.001}$,  Rh$_{0.999}\,M^{5d}_{0.001}$ and  Nb$_{0.999}\,M^{5d}_{0.001}$, 
composed of the $4d$ impurities in $5d$-metal host and the other way
around, of $5d$ impurities in $4d$-metal host. 
The corresponding OHC and SHC are plotted in Fig.\ \ref{Me1_vs_Me2_OHE_SHE}.
First of all, one can see for all doped materials, that the absolute
values of the OHC are smaller than those of the SHE. The OHC obtained
for Ir$_{0.999}\,M^{4d}_{0.001}$ and for Rh$_{0.999}\,M^{5d}_{0.001}$
have opposite sign. The same property is found also for
the SHC. Such a behavior has been observed previously for SHC 
\cite{Low10} and was attributed to the relative strength of the SOC
associated with the host and impurity atoms. Some further features concerning
the impact of the SOC on the OHC and SHC will be discussed also below.

Another prominent feature of the OHC and SHC shown in
Figs.\ \ref{Me1_vs_Me2_OHE_SHE}, is a strong enhancement of the OHC and 
the SHC in the case, when the impurity atoms have the number of valence
electrons close to that of the host atoms. As one can see,
it is a consequence of the enhancement of the extrinsic OHC and SHC 
representing the leading contributions to the total OHC and SHC.

\begin{figure}[t]
  \begin{center}
 \includegraphics[angle=270,width=0.45\linewidth,clip]{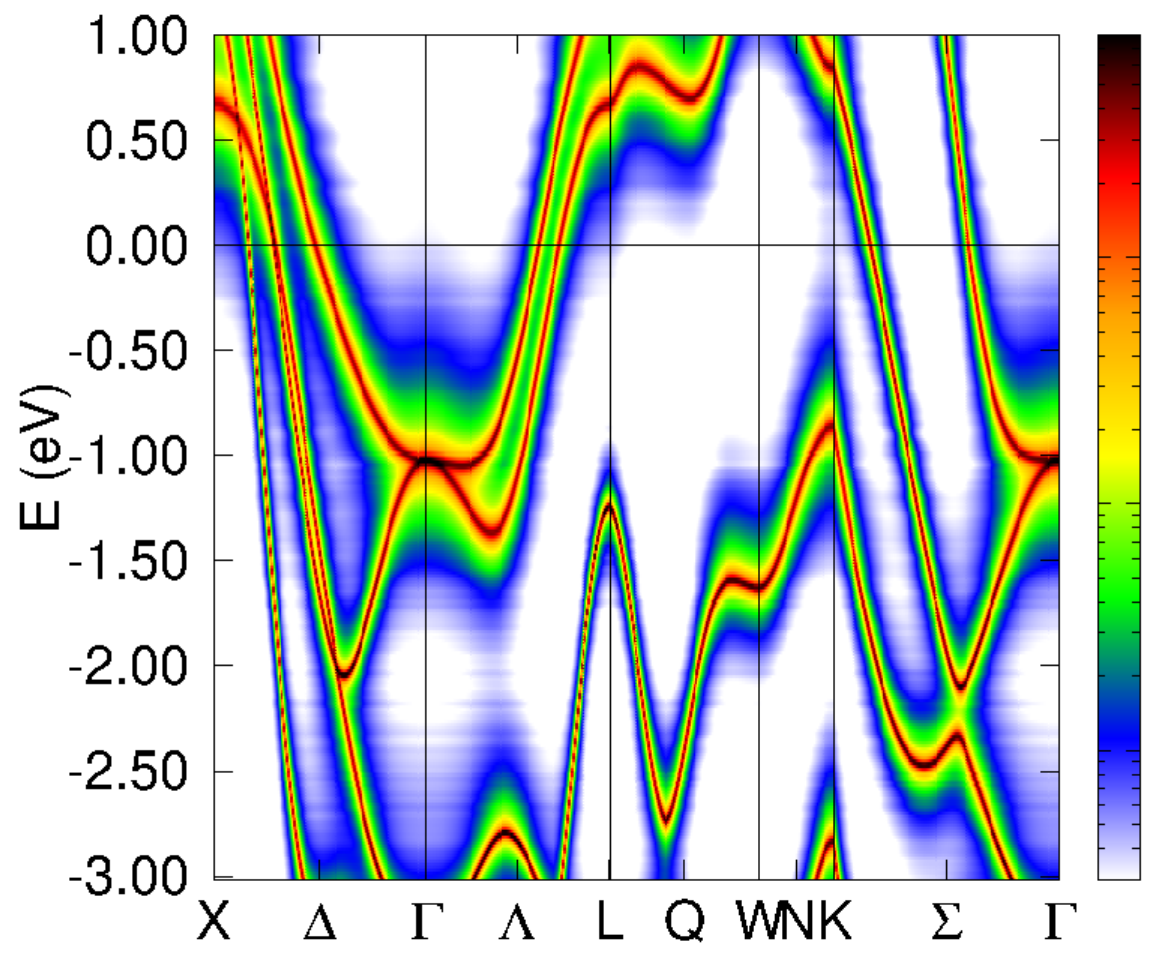}
 \includegraphics[angle=270,width=0.45\linewidth,clip]{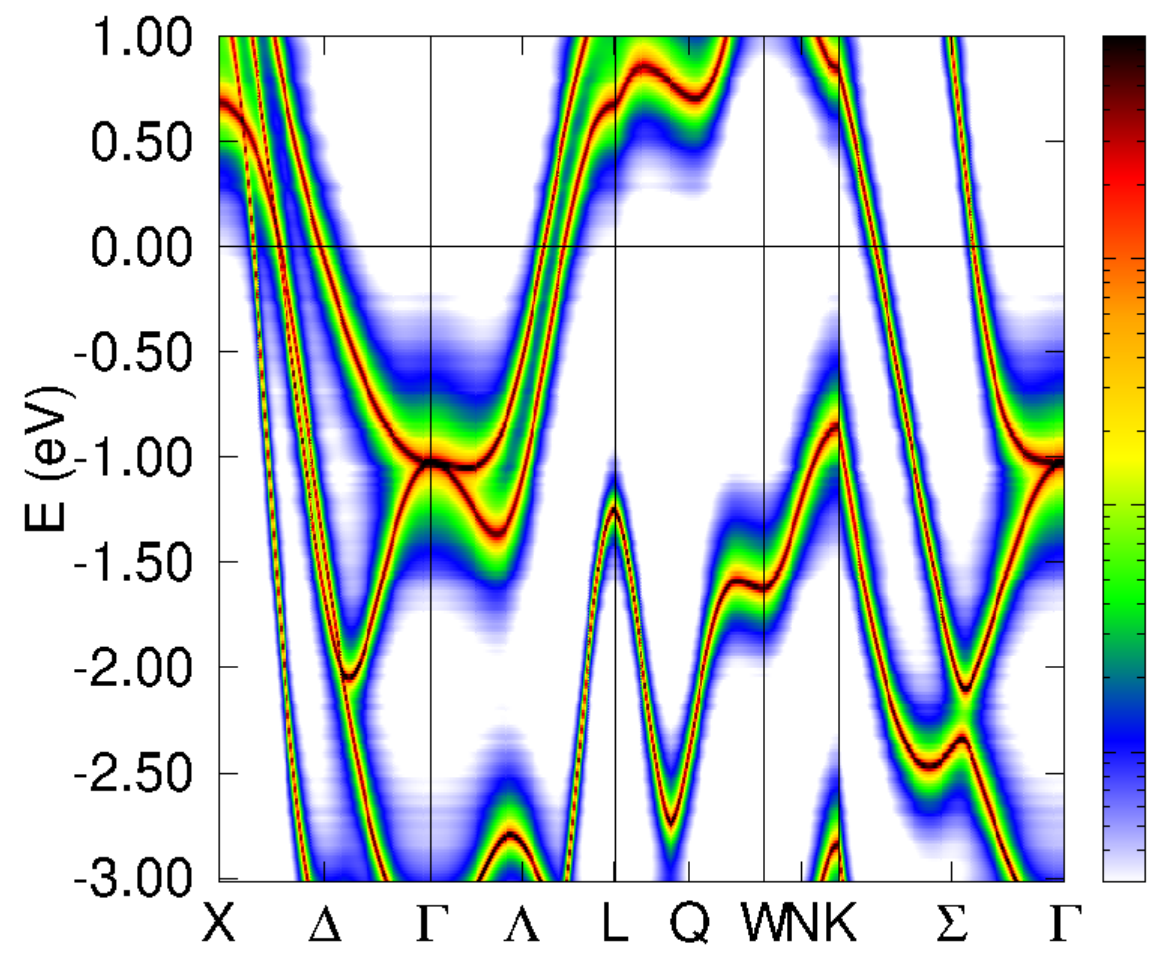}\\
\;\;\;  (a) \;\;\;\;\;\; \;\;\;\;\;\;\;\; \;\;\;\;\;\;\;\; \;\;\;\;\; \; \;   (b)\\
 \includegraphics[angle=270,width=0.45\linewidth,clip]{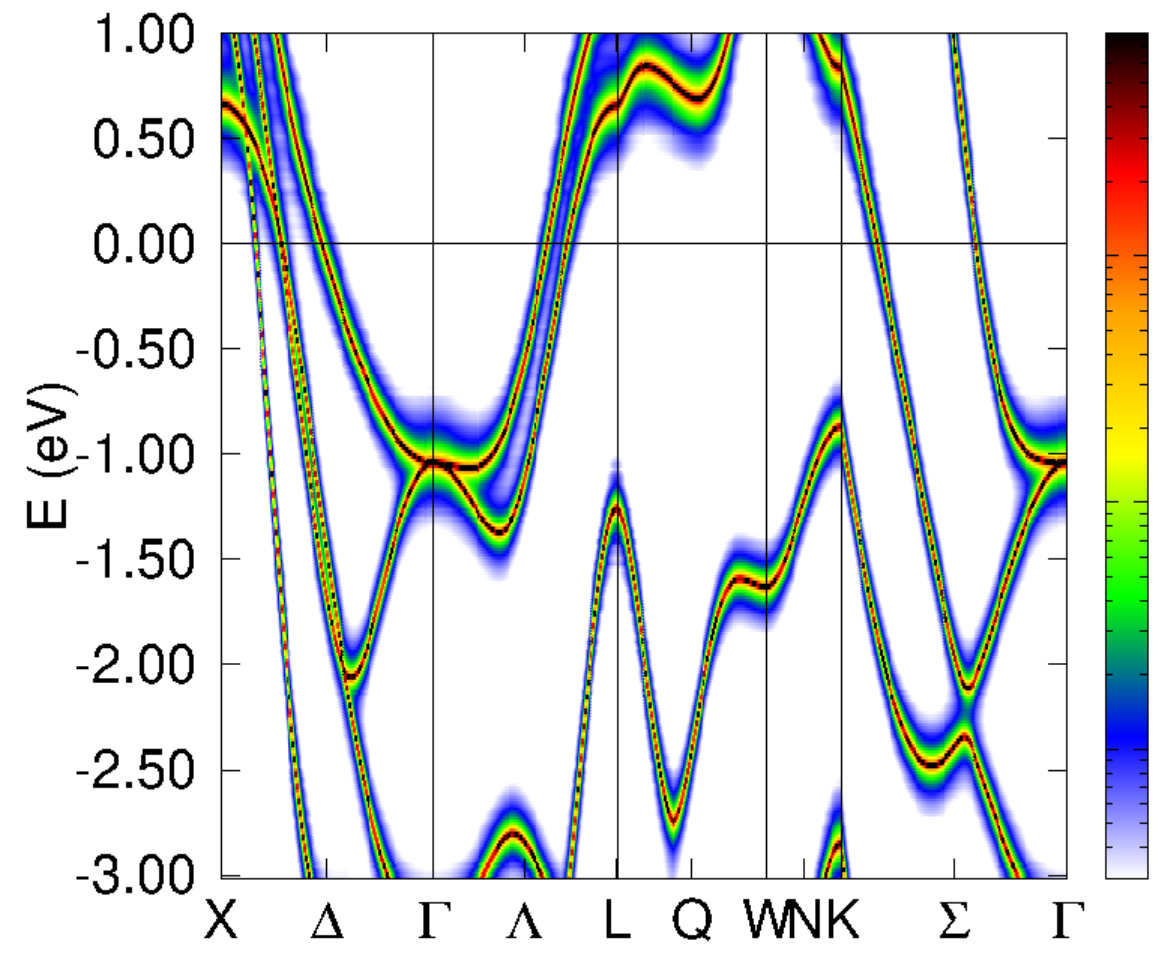}
 \includegraphics[angle=270,width=0.45\linewidth,clip]{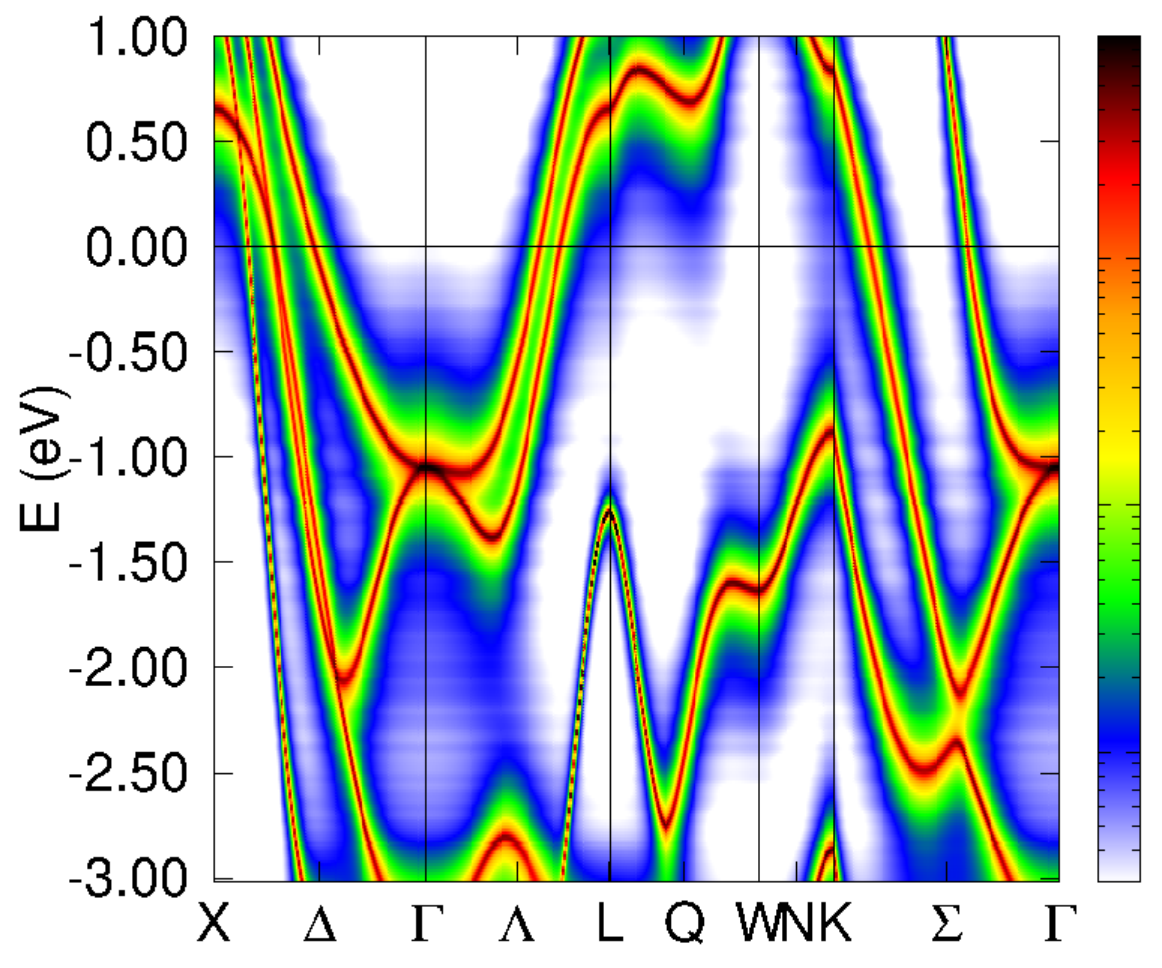}\\
\;\;\;  (c) \;\;\;\;\;\; \;\;\;\;\;\;\;\; \;\;\;\;\;\;\;\; \;\;\;\;\; \; \;   (d)\\
  \caption{\label{Fig_BSF_Ir_IrRh} BSF $A(E,\vec{k})$ for  Ir$_{0.99}$Mo$_{0.01}$  (a)
    Ir$_{0.99}$Tc$_{0.01}$ (b) Ir$_{0.99}$Rh$_{0.01}$ (c) and Ir$_{0.99}$Ag$_{0.01}$ (d).
}
\end{center}
\end{figure}

To figure out the origin of such a behavior, we can consider the
impact of different impurities on the electronic structure of the doped
materials. For this purpose, Fig.\ \ref{Fig_BSF_Ir_IrRh} shows the Bloch spectral
function (BSF) $A(\vec{k},E)$ for selected Ir$_{0.99}M^{4d}_{0.01}$
compounds, plotted along the high-symmetry directions of the Brillouin
zone (BZ). 
One can clearly see a smearing of the energy bands due to disorder, which
is rather moderate in the case of Ir$_{0.99}$Rh$_{0.01}$, i.e., for the system
with the host and impurity elements having the same number of valence
electrons. The most prominent smearing is observed for
Ir$_{0.99}Mo_{0.01}$ and Ir$_{0.99}$Ag$_{0.01}$ for which the OHC and SHC
are the smallest among those shown in  Fig.\ \ref{Me1_vs_Me2_OHE_SHE}. 

\begin{figure}[t]
  \begin{center}
 \includegraphics[angle=0,width=0.8\linewidth,clip]{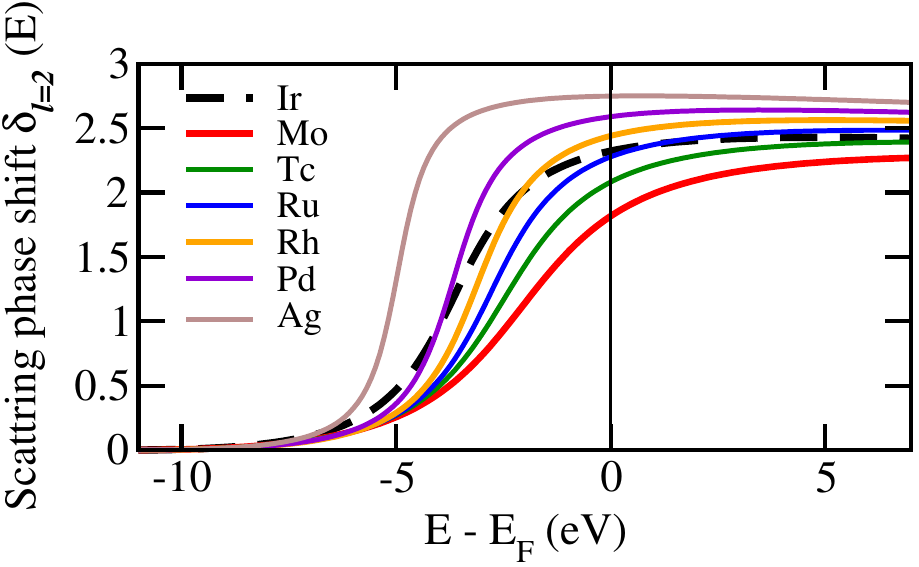}
  \caption{\label{Fig_Phase_Ir-Me} Scattering phase shift
    $\delta_{l=2}(E)$ for Ir$_{0.99}M_{0.01}^{4d}$. 
}
\end{center}
\end{figure}

Thus, one can clearly see a correlation between the values of the OHC and SHC and
the smearing of the electron energy bands (i.e. inverse lifetime)
caused by different types of impurities.
 We will discuss it in terms of the energy of the d-band 
  scattering resonance $E^M_d$ of the host and impurity atoms, varying
  across the transition metal 
  period (see, e.g. \cite{SWF71}), which may be associated with the
  position of the center of $d$-band in metal. With this, the scattering
  strength $V_s$ may be characterized by the distance between the
  centers of valence-electron 
  $d$-bands of the host and impurity atoms, $V_s \sim \Delta_d$, that leads to estimate
  for the lifetime of the electron states $\tau \propto  \Delta_d^{-2}$
  (while  $\tau^{-1} \propto \langle V_s^2 \rangle$ \cite{CB01}).  
  This implies, that the weak smearing of the $d$-states,  
  shown Fig.\ \ref{Fig_BSF_Ir_IrRh},  
  should be small in the case of a small $\Delta_d$, and should
  increase together with increasing $\Delta_d$ value,
  as it is inversely proportional to their lifetime $\tau$.
  To visualize this trend for Ir$_{0.99}\,M^{4d}_{0.01}$ materials,
  Fig.\ \ref{Fig_Phase_Ir-Me} shows the scattering phase shift  
  $\delta_{l=2}(E)$ showing the $d$-resonances
  of the impurity and host atoms. 
  Note that for simplicity reason the results are given for the
  non-relativistic case.
  The dashed line represents the phase shift $\delta_{l=2}^{Ir}(E)$ for the
  host atom. As one can see, the position of the $4d$-resonances of the
impurity atoms changes when the impurity $M$ is varied along the $4d$ period,
crossing the resonance energy for the host Ir atom.
The results shown in Fig.\ \ref{Fig_Phase_Ir-Me} are in line with the
BSF plotted in Fig.\ \ref{Fig_BSF_Ir_IrRh}.
When the resonances are close to each other one can
see the weakest scattering amplitude, e.g. in the case of
Ir$_{0.99}$Rh$_{0.01}$ in contrast to Ir$_{0.99}$Ag$_{0.01}$  with
$\Delta_d \approx 0.2$ eV and $\Delta_d \approx 1.6$ eV, respectively. 
Furthermore, using the estimate $\sigma_{\nu\nu} \propto
\Delta_d^{-2}$ as discussed by Cr\'epieux and Bruno \cite{CB01},
one can get an idea on the impact of the type of impurity atom
on the transport properties.  
Figs. \ref{Me1_vs_Me2_OHE_SHE}(c) and (d) show the electrical
conductivity for the Ir$_{0.999}\,M^{4d}_{0.001}$, 
Rh$_{0.999}\,M^{5d}_{0.001}$ and
Nb$_{0.999}\,M^{5d}_{0.001}$ systems. In line with our discussions 
above, the maximum of the conductivity occurs in the case,
when the distance $\Delta_d$  between the $4d$ and $5d$ resonances
is minimal. On the other hand, the skew-scattering conductivity follows
the relationship $\sigma_{\rm OH/SH}^{\rm skew} \propto \langle V_s
\rangle^3 / \langle V_s^2 \rangle^2 $ (see 
Ref.\ [\onlinecite{CB01}]). As a consequence, this leads to
$\sigma_{\rm OH/SH}^{\rm skew} \propto \Delta_d^{-1}$, i.e. to the
element dependent enhancement of the OHC and SHC occurring in parallel
with the enhancement of the electrical conductivity.

\subsubsection{Temperature effect in doped systems}

\begin{figure}
 \begin{center}
 \includegraphics[angle=0,width=0.43\linewidth,clip]{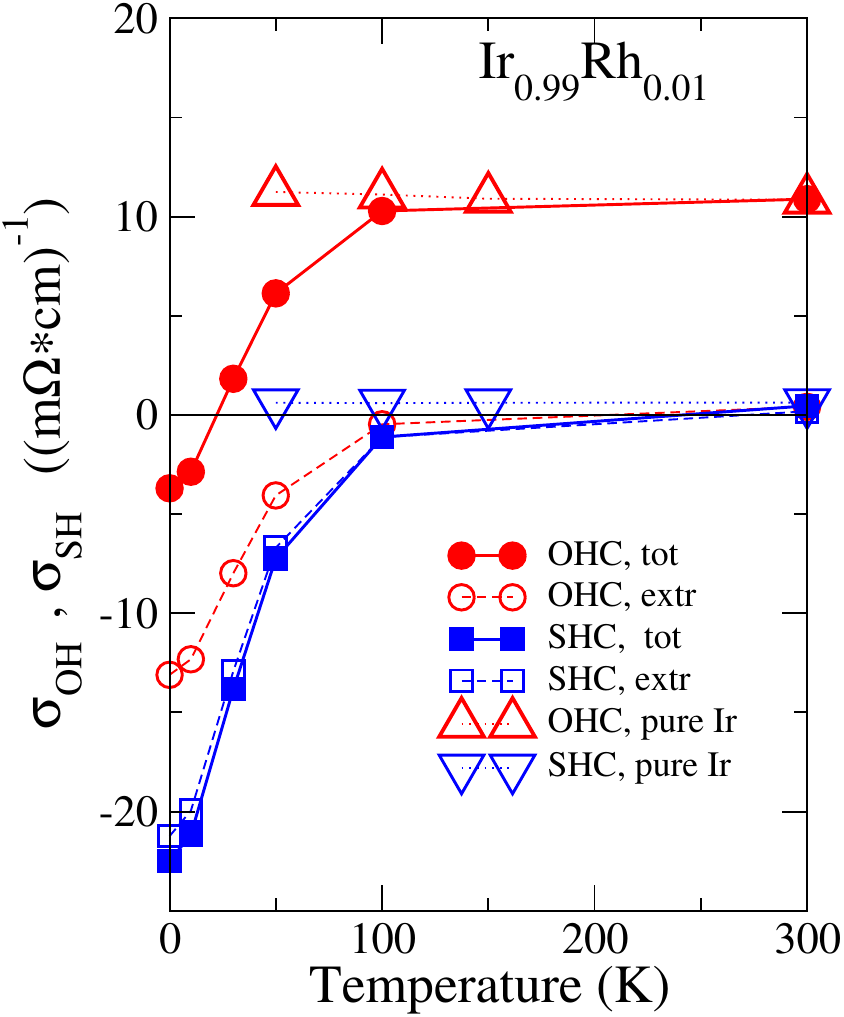}\,(a)
 \includegraphics[angle=0,width=0.43\linewidth,clip]{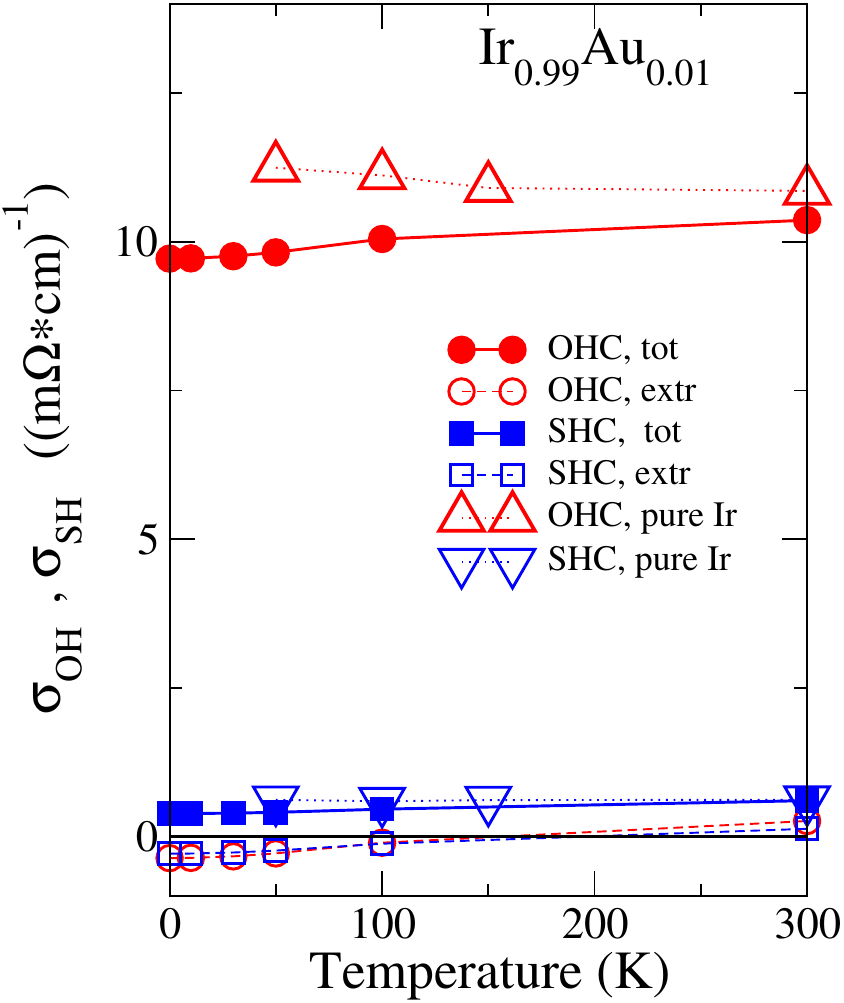}\,(b)
  \caption{\label{Me1-Me2_OHE_SHE-vs-T} OH (circles) and SH (squares)
    conductivities for  Ir$_{0.99}$Rh$_{0.01}$ (a) and
    Ir$_{0.99}$Au$_{0.01}$ (b) plotted as a function of 
  temperature. Full symbols represent total values calculated
  accounting for vertex corrections, why open symbols show only the
  extrinsic contribution to the OHC and SHC. Triangles represent the
  results on  the OHC and SHC for pure Ir.}
\end{center}
\end{figure}

Next, we will discuss the impact of temperature induced lattice vibrations
on the OHE and SHE for doped
materials. Fig.\ \ref{Me1-Me2_OHE_SHE-vs-T} demonstrates the OHC
(circles) and SHC 
(squares) for Ir$_{0.99}$Rh$_{0.01}$ (a) and Ir$_{0.99}$Au$_{0.01}$ (b),
plotted as a function of temperature, 
 and compared with the results for pure Ir.
The most pronounced difference between the results for doped and
undoped systems occurs at low temperature. 
This difference gradually vanishes in the high-temperature regime,
when the OHC and SHC for doped and undoped systems approach the values
mainly given by the intrinsic contributions.
 The strong temperature dependence of the OHC and SHC for Ir$_{0.99}$Rh$_{0.01}$
  at low temperature (see Fig.\ \ref{Me1-Me2_OHE_SHE-vs-T} (a)) stems
  from the extrinsic skew-scattering contributions that quickly decrease at
  raising temperature. As $\sigma_{\rm OH/SH}^{\rm skew} \propto \langle
  V_s \rangle^3 / \langle V_s^2 \rangle^2 $, at finite temperature the
  scattering potential can be split into 'atomic' and
  'electron-phonon' parts. While the nominator $\langle V_s \rangle^3
  \approx \langle V_a \rangle^3$, as $ \langle V_{e-ph} \rangle = 0$
  \cite{CB01}, the denominator accounts for both types of scattering
  contributions, i.e. $ \langle 
  (V_{a} + V_{e-ph})^2 \rangle$, increasing with temperature due to
 an increasing amplitude of the lattice vibrations.

  Thus, in the case of Ir$_{0.99}$Rh$_{0.01}$, a large value of
  the skew scattering conductivity at $T = 0$ K, $\sigma_{\rm OH/SH}^{\rm skew} \propto
  \frac{1}{x}\frac{1}{\Delta_d}$ \cite{CB01} with $\Delta_d \approx 0.2$
  eV, decreases at finite temperature due to electron-phonon scattering
  (up to $\langle V_{e-ph}^2 \rangle^{1/2} \sim 0.1$ eV) according to
  $\sigma_{\rm OH/SH}^{\rm skew} \propto \langle V_a \rangle^3 / \langle
  (V_a + V_{e-ph}(T))^2 \rangle^2 $. 
  In the case of Ir$_{0.99}$Au$_{0.01}$, however (see
  Fig.\ \ref{Me1-Me2_OHE_SHE-vs-T}(b)) $\Delta_d >> V_{e-ph}$ as
  $\Delta_d \approx 1.6$ eV, leading to a weak dependence on temperature
  according to 
  $\sigma^{\rm skew}_{\rm OH/SH} \propto \frac{1}{x}\frac{1}{\Delta_d}
 (1 - O(\langle V_{el-ph}^2\rangle / \langle V_a^2\rangle)) $.
  At high temperature  $\sigma_{\rm OH}$ and $\sigma_{\rm SH}$ approach
  the intrinsic OHC and SHC, as it was seen also for Ir$_{0.99}$Rh$_{0.01}$.

\subsubsection{SOC effect}

\begin{figure}
 \begin{center}
 \includegraphics[angle=0,width=0.43\linewidth,clip]{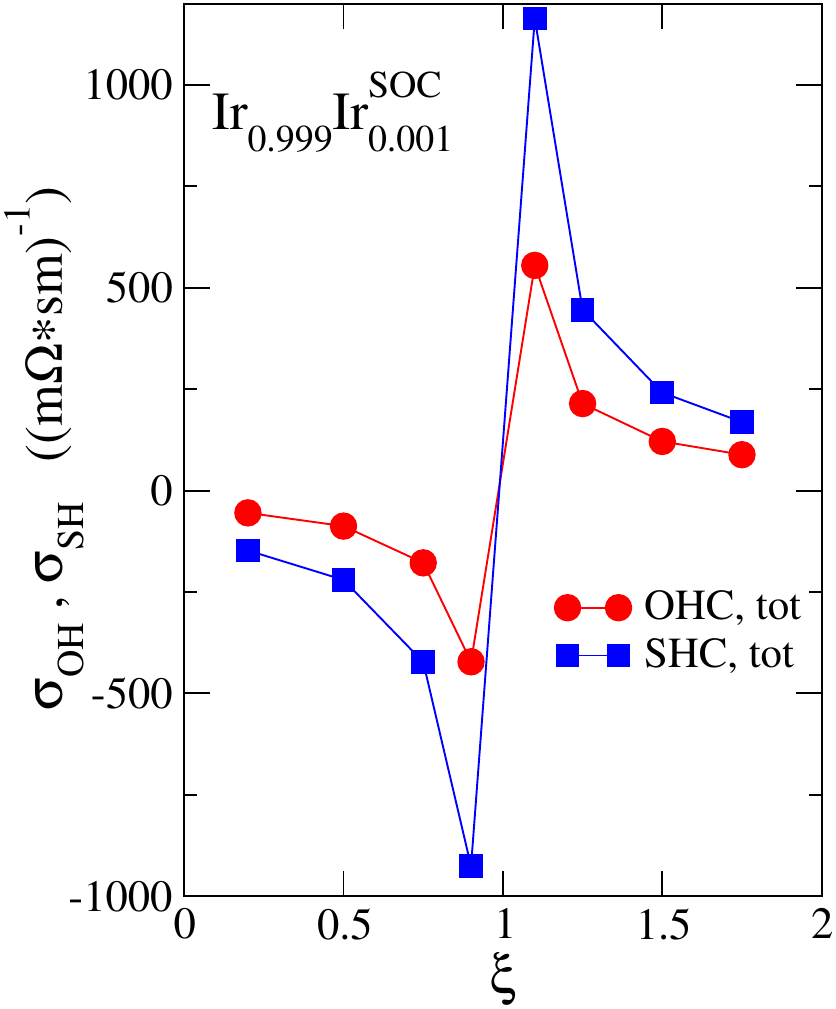}\,(a)
 \includegraphics[angle=0,width=0.43\linewidth,clip]{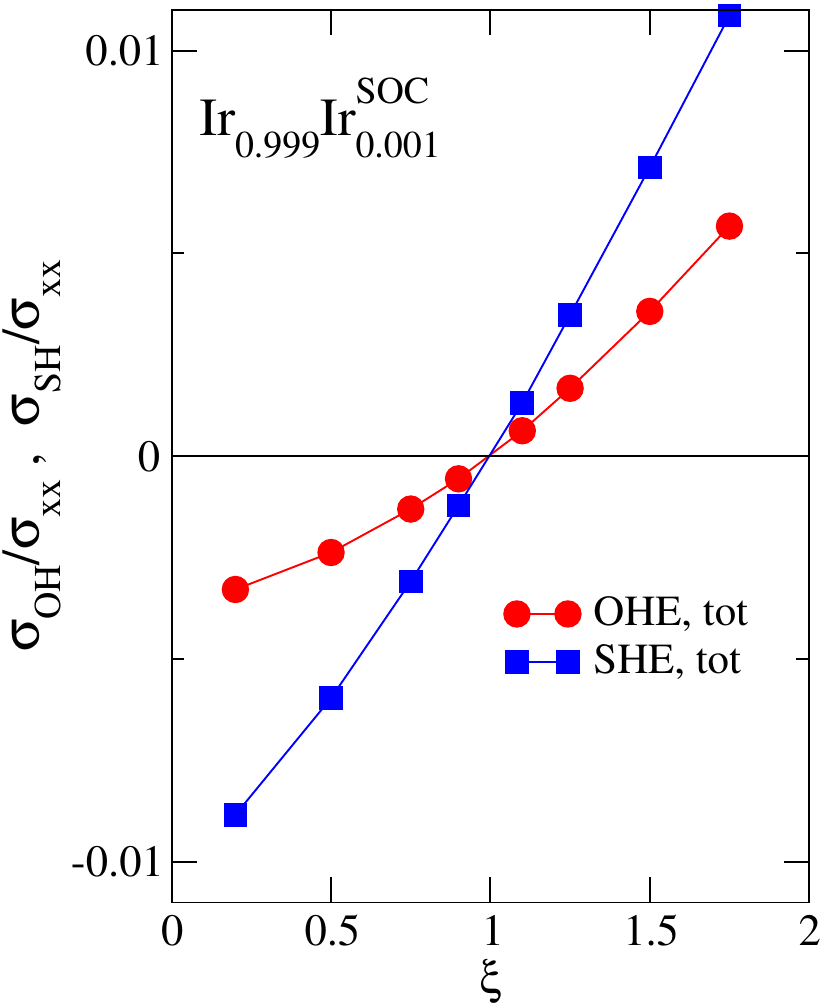}\,(b)
  \caption{\label{Ir-Ir0.001_OHE_SHE-vs-SOC} The OH and SH conductivities
    for  Ir$_{0.999}$Ir$_{0.001}^{SOC}$, i.e. for Ir impurities in Ir
    with the speed of light scaled by a factor $\xi^{-1}$: (a)
    absolute values of OHC and SHC and (b) OHC and SHC are divided by
    electrical conductivity, i.e.  $\sigma_{\rm OH}/\sigma_{xx}$ and $\sigma_{\rm SH}/\sigma_{xx}$. }
\end{center}
\end{figure}

Finally, we demonstrate the impact on the OHC and SHC of only 
SOC-driven scattering events. For this, we consider a model system,
Ir$_{0.999}$Ir$_{0.001}$, composed of the
same type of atoms used for host as well as for impurity.
In the case of unchanged SOC in the system, it is identical to pure Ir.
However, using the parameter $\xi = c_0^{2}/c^{2}$ to scale speed of
light at the impurity sites, one can create SOC-driven
scattering potential given as follows
\begin{equation}
  V_{imp}^{SOC} = (\xi - 1) \frac{1}{2m^2c_0^2r} \frac{\partial V}{\partial
    r}(\hat{\vec{s}}\cdot\hat{\vec{l}})\; ,
\end{equation}
with $\hat{\vec{s}} =\frac{\hbar}{2} \vec{\sigma}$ and
$\cdot\hat{\vec{l}}$ being the operators of spin and orbital angular momentum.
Fig.\ \ref{Ir-Ir0.001_OHE_SHE-vs-SOC} (a) represent the results on the OHC
and SHC for such a system, plotted as a function of scaling
factor $\xi$. One can see a divergence of both the OH and SH
conductivities, $\sigma_{\rm OH}$ and $\sigma_{\rm SH}$ in the limit of
$\xi - 1 \to \pm 0$, with opposite signs for $\xi
< 1$ and $\xi > 1$, arising due to extrinsic contributions.
This behavior is a consequence of vanishing
scattering potential in the case of  $\xi = 1$ leading also to a
divergence of the electrical conductivity.
Assuming that the behavior of both extrinsic OH and SH conductivities are
determined by skew scattering and are proportional to electrical
resistivity, the ratios $\sigma_{\rm OH}/\sigma_{xx}$ and
$\sigma_{\rm SH}/\sigma_{xx}$ should be smooth functions of scaling factor
$\xi$, that can be seen in Fig.\ \ref{Ir-Ir0.001_OHE_SHE-vs-SOC}
(b). These results also ensure a common origin of  the extrinsic OH and
SH conductivities. 
Also, these results demonstrate a dependence of the sign of these
quantities on the relative values of SOC corresponding to the host atoms
and to impurities.

\section{Summary}

To summarize, the OHC calculated for non-magnetic $3d$-, $4d$-
and $5d$-metals, elemental and doped, is compared with corresponding
results on the SHC.
 A strong difference has been found for the temperature dependence
 of the OHC and SHC for undoped and doped
 systems. For elemental systems at finite temperature,
 the electron-phonon skew scattering has a negligible impact on the
 extrinsic OHC and SHC. As a consequence, the OHC and SHC and their
 temperature dependent variation is determined mainly by the intrinsic
 contributions. Moreover, different temperature dependent behavior of
 the intrinsic SOC-independent OHC and SOC-driven SHC indicates
 non-trivial relationship between these quantities. 
 In doped materials, the extrinsic contribution to the OHC and SHC is
 dominating at low temperature. However, a strong  decrease is found for
 higher temperatures due to the increasing impact  of the
 electron-phonon scattering.  

 \begin{acknowledgments}
  The authors thank Erna Delczeg for bringing our attention to the
  subject studied in the work, as well as for useful discussions. 
 \end{acknowledgments}


\appendix
\section{Kubo-Bastin equation}

The so-called Bastin formula for the response quantity $\sigma$ is given
by the following equation\cite{BLBN71} 
\begin{widetext}
\begin{equation}
\label{eq:Ba}
\sigma_{\mu\nu} =  \frac{i \hbar}{V} \int_{-\infty}^{\infty} dE f(E) {\rm Tr}\Bigg< \hat{J}_{\mu} \frac{d G^+(E)}{d E}\hat{j}_{\nu}\delta(E -\hat{H}) -  \hat{J}_{\mu}\delta(E -\hat{H})  \hat{j}_{\nu}  \frac{d G^-(E)}{d E} \Bigg>_c \;
\end{equation}  \\ \medskip \\
with $\nu = (x,y,z)$ denoting  Cartesian  coordinates, $f(E)$ the
Fermi-Dirac distribution function and $G^{\pm}(E)= (E-\hat{H}\pm
i\delta)^{-1}$ the retarded and advanced Green’s function operators.
The equation describes the response of  an observable  represented  by  an operator
$\hat{J}_{\mu}$ to the perturbation represented by the operator
$\hat{j}_{\nu}$.
The brackets $\langle ... \rangle$ denotes a configurational average in
the presence of any type of disorder.
For the case $T = 0$ K, this expression can be transformed to the following form\cite{KCE15}:

\begin{eqnarray}
  \label{eq:CP-rewrite-4}
  \sigma^{\xi}_{\mu\nu}
  &=&
    \frac{\hbar}{4\pi\Omega} \Bigg[
    {\rm Trace}
    \left<
      \hat{J}^{\xi}_{\mu}(\hat{G}^{+}-\hat{G}^{-}) \hat{j}_{\nu} \hat{G}^{-}
      -\hat{J}^{\xi}_{\mu} \hat{G}^{+} \hat{j}_{\nu}(\hat{G}^{+}-\hat{G}^{-})
    \right>\\
   &&   + \int_{-\infty}^{E_F}d\varepsilon
    {\rm Trace}
    \left<
       \hat{J}^{\xi}_{\mu} \hat{G}^{+}\hat{j}_{\nu}
        \frac{d\hat{G}^{+}}{d\varepsilon}
        -\hat{J}^{\xi}_{\mu}\frac{d\hat{G}^{+}}{d\varepsilon} \hat{j}_{\nu} \hat{G}^{+}\right.
        -  \left.\left(
          \hat{J}^{\xi}_{\mu} \hat{G}^{-}\hat{j}_{\nu}\frac{d\hat{G}^{-}}{d\varepsilon}
          -\hat{J}^{\xi}_{\mu}\frac{d\hat{G}^{-}}{d\varepsilon} \hat{j}_{\nu} \hat{G}^{-}
        \right)\right> \Bigg] =  \sigma_{\mu\nu}^{I\xi}+\sigma_{\mu\nu}^{II\xi}
\end{eqnarray}
\end{widetext}
  
The first term is evaluated at the Fermi energy $E_{\rm                       
F}$ and contains contributions only from the Fermi surface and is
referred to as Fermi-surface term. The second term implies 
an integration over all occupied states and is called Fermi-sea
term.
Within the multiple-scattering or KKR formalism, the real-space Green’s
function $G(\vec{r},\vec{r}\,',E)$ is given by the expression
\cite{EBKM16}:  
\begin{eqnarray}
G(\vec{r},\vec{r}\,',E) & = &
\sum_{\Lambda_1\Lambda_2} 
Z^{n}_{\Lambda_1}(\vec{r},E)
                              {\tau}^{n n'}_{\Lambda_1\Lambda_2}(E)
Z^{n' \times}_{\Lambda_2}(\vec{r}\,',E)
 \nonumber \\
 & & 
-  \sum_{\Lambda_1} \Big[ 
Z^{n}_{\Lambda_1}(\vec{r},E) J^{n \times}_{\Lambda_1}(\vec{r}\,',E)
\Theta(r'-r)  \nonumber 
\\
 & & \qquad\quad 
J^{n}_{\Lambda_1}(\vec{r},E) Z^{n \times}_{\Lambda_1}(\vec{r}\,',E) \Theta(r-r')
\Big] \delta_{nn'} \; . \nonumber \\
\label{Eq_KKR-GF}
\end{eqnarray}
Here $\vec{r},\vec{r}'$ refer to site $n$ and $n'$, respectively,
${\tau}^{nn'}_{ \Lambda  \Lambda'} (E)$ is the so-called scattering path
operator that transfers an electronic wave coming in at 
site $ n' $ into a wave going out from site $ n $ with
all possible intermediate scattering events accounted for. 
The four-component wave functions $Z^{n}_{\Lambda}(\vec r,E)$ 
($J^{n}_{\Lambda}(\vec r,E)$) are regular (irregular)
solutions to the single-site Dirac equation \cite{Tam92} with the
Hamiltonian set up within the framework of relativistic spin-density
functional theory \cite{MV79,ED11}: 
\begin{eqnarray}
{\cal H}_{\rm D}  & =&
- i c \vec{ \alpha} \cdot \vec \nabla  + \frac{1}{2} \, c^{2} ({\beta} - 1) + V(\vec r)  + \beta \vec{\sigma}\cdot {\vec B}_{xc}(\vec r)
\; . \nonumber \\
\label{Hamiltonian}           
\end{eqnarray}
These functions are labeled by the combined quantum numbers $\Lambda$
($\Lambda = (\kappa,\mu)$), with 
$\kappa$ and $\mu$  being the spin-orbit and magnetic quantum numbers
\cite{Ros61}. The superscript $\times$ indicates the left hand
side solution of the Dirac equation.
The operators  $ {\alpha}_i $ and $ \beta $ in the Hamiltonian in Eq.\
(\ref{Hamiltonian}) are the standard Dirac matrices \cite{Ros61}  while
 $\bar V(\vec {r}) $ and ${\vec B}_{xc}(\vec r)$ are the spin
 independent and dependent parts of the electronic potential
 \cite{Ros61,EBKM16}. 

In the case of disordered systems, the configurational
average in the expression for the response tensor in
Eq.\ [\ref{eq:CP-rewrite-4}], $\langle J G^{\pm} j G^{\pm} \rangle
\approx \langle J G^{\pm} \rangle \langle j G^{\pm}\rangle$, is
performed making use of a configuratinal average of the single-particle Green
function in the spirit of the coherent potential approximation (CPA)
alloy theory, and accounting for the
vertex corrections referring to two-particle quantities in the
expression for the conductivity. 
Formally a corresponding response tensor element may be written as
follows \cite{Low10} (omitting the angular momentum indices): 
\begin{eqnarray}
\sigma_{\mu\nu}^{\rm VC} &\propto& \, Tr {J}_{\mu}(z_2,z_1)\,
                               (1-\chi\,w)^{-1}\chi\,
                                {j}_{\nu}(z_1,z_2) \,.
\label{eq:sigma_VC} \\
\sigma_{\mu\nu}^{\rm NVC} &\propto& \, Tr  {J}_{\mu}(z_2,z_1)\, \chi
                                  \, {j}_{\nu}(z_2,z_1)\,
\label{eq:sigma_NVC} 
\end{eqnarray}
with the term $(1-\chi\,w)^{-1}$ accounting for vertex corrections
\cite{But85}. Accordingly, VC and NVC denotes the response quantity
including and excluding , respectively, the vertex corrections.
The auxiliary quantity $\chi$ in Eqs.\ (\ref{eq:sigma_VC}) and
(\ref{eq:sigma_NVC}) is given by 
\begin{eqnarray} 
\nonumber
 && \chi_{\Lambda_1 \Lambda_2 \Lambda_3 \Lambda_4}(z_1,z_2) \nonumber \\  & = &
\frac{1}{\Omega_{\rm BZ}}\int_{\rm BZ} d^3k \,\tau_{{\rm CPA},\Lambda_1
  \Lambda_2}({\bf k},z_1)\tau_{{\rm CPA},\Lambda_3 \Lambda_4}({\bf
  k},z_2) \nonumber \\
&& - \tau^{00}_{\,{\rm CPA},\Lambda_1 \Lambda_2}(z_1)\tau^{00}_{\,{\rm CPA},\Lambda_3 \Lambda_4}(z_2)  \; ,
\end{eqnarray}
and the interaction term $w$ is defined as
\begin{align}
\label{eq:x_cpa2}
w_{\Lambda_1 \Lambda_2 \Lambda_3 \Lambda_4}(z_1,z_2) &= & \sum_{\alpha} x_{\alpha} x_{\Lambda_1 \Lambda_2}^{\alpha}(z_1)x_{\Lambda_3 \Lambda_4}^{\alpha}(z_2)\; 
\end{align}
with
\begin{align}
\label{eq:x_cpa1}
\underline{x}^{\alpha}(z) & = & \left\{
\left[\underline{t}_{\alpha}^{-1}(z) - \underline{t}_{\rm CPA}^{-1}(z)
  \right]^{-1} +  \underline{\tau}_{\,\rm CPA}^{00}(z) \right\} \; ,
\end{align}
where $\underline{t}_{\,\alpha}$ and $ \underline{t}_{\,\rm
CPA}$ represent the single site $t$-matrix for atom type $\alpha$ and
for the CPA medium, respectively, and $\underline{\tau}_{\,{\rm CPA}}$
denotes the CPA averaged $\tau$ matrix. More details can be found in
Ref.\ [\onlinecite{KCE15}].

Following Butler’s scheme \cite{But85},
a response tensor may be spitted into a site-diagonal  
term ${\sigma}^{0}_{\mu\nu} \propto \langle J^{0\mu}\tau^{00}
j^{0\nu}\tau^{00}\rangle$ and site-off-diagonal term 
${\sigma}^{1}_{\mu\nu} \propto \langle J^{n\mu}\tau^{nm}
j^{m\nu}\tau^{mn}\rangle$ \cite{But85,Low10}, such that the former term
is purely coherent.
 Furthermore, according to the findings by Turek et al. \cite{TKD19}
concerning the SHC, the vertex corrections to the site-off-diagonal
Fermi sea term is negligible.  
This implies at the end, that the extrinsic SHC is dominated by the
Fermi surface contribution, ${\sigma}^{1}_{\mu\nu} (E_F)$.


%

\end{document}